\documentclass[3p,authoryear]{elsarticle}
\usepackage[utf8]{inputenc}
\usepackage{amssymb}
\usepackage{amsmath}
\usepackage{amsthm}
\usepackage{mathtools}
\usepackage{algorithm}
\usepackage{algpseudocode}
\usepackage{booktabs}

\usepackage{mathtools}
\usepackage{booktabs}
\usepackage{multirow}
\usepackage{graphicx}
\usepackage[table,xcdraw]{xcolor}
\title{Identifying Critical Fleet Sizes Using a Novel Agent-Based Modelling Framework for Autonomous Ride-Sourcing}

\author[1]{Renos Karamanis\corref{cor1}}
\ead{renos.karamanis10@imperial.ac.uk}
\author[1]{He-in Cheong}
\author[2]{Simon Hu}
\author[1]{Marc Stettler}
\author[1]{Panagiotis Angeloudis}

\cortext[cor1]{Corresponding author: renos.karamanis10@imperial.ac.uk (R. Karamanis), Skempton Buidling, SW7 2BB, UK}
\cortext[cor2]{This work was supported by the Engineering and Physical Sciences Research Council (EPSRC).}
\address[1]{Centre for Transport Studies, Imperial College London}
\address[2]{International Campus, Zhejiang University}

\date{November 2020}
\begin{document}

\begin{abstract}
    Ride-sourcing platforms enable an on-demand shared transport service by solving decision problems often related to customer matching, pricing and vehicle routing. These problems have been frequently represented using aggregated mathematical models and solved via algorithmic approaches designed by researchers. The increasing complexity of ride-sourcing environments compromises the accuracy of aggregated methods. It, therefore, signals the need for alternative practices such as agent-based models which capture the level of complex dynamics in ride-sourcing systems. The use of these agent-based models to simulate ride-sourcing fleets has been a focal point of many studies; however, this occurred in the absence of a prescribed approach on how to build the models to mimic fleet operations realistically. To bridge this research gap, we provide a framework for building bespoke agent-based models for ride-sourcing fleets, derived from the fundamentals of agent-based modelling theory. We also introduce a model building sequence of the different modules necessary to structure a simulator based on our framework. To showcase the strength of our framework, we use it to tackle the highly non-linear problem of minimum fleet size estimation for autonomous ride-sourcing fleets. We do so by investigating the relationship of system parameters based on queuing theory principles and by deriving and validating a novel model for pickup wait times. By modelling the ride-sourcing fleet function in the urban areas of Manhattan, San Francisco, Paris and Barcelona, we find that ride-sourcing fleets operate queues with zero assignment times above the critical fleet size. We also show that pickup wait times have a pivotal role in the estimation of the minimum fleet size in ride-sourcing operations, with agent-based modelling to be a more reliable route for their identification given the system parameters. Our proposed agent-based modelling framework can be used to investigate complex problems and streamline reproducibility of research in ride-sourcing systems.
\end{abstract}


\maketitle

\section{Introduction} \label{sec: Intro}
The widespread penetration of smartphone technology and the availability of high-speed internet over the cellular network facilitated the evolution of urban on-demand transportation services from traditional taxis to ride-sourcing platforms. These platforms, otherwise referred to as Transportation Network Companies (TNCs), operate as technology firms, by efficiently connecting incoming ride requests with available drivers in the network. In the last few years, TNCs managed to acquire large portions of the on-demand transportation market in cities due to their operational efficiency \citep{Schneider2020}. The digitization of on-demand transportation and the anticipated deployment of Autonomous Vehicles (AVs) in TNC services to exploit economies of scale, motivated a plethora of AV TNC related studies in recent years \citep{Wang2019}.

 The introduction of AVs in TNC services could transform TNC platforms into fleet operators \citep{qiu2018dynamic}. Consequently, the typical list of decisions for a TNC, which includes assignment and pricing of private or shared rides, would be extended with fleet management functions, such as empty vehicle redistribution to areas of excess demand, as well as electric charging and parking operations. As TNCs are interested in running efficient services, the multitude of their decision problems is governed by mathematical models, usually in the form of optimisation problems, either structured as aggregated representations of the system or specific sub-problem instances in the fleet operation \citep{narayanan2020shared}.
 
 Developers of TNC platforms are usually interested in system-wide Key Performance Indicators (KPIs) to quantify the impact of their models. These KPIs might not always be identifiable from the mathematical models. Some examples of such metrics include the velocities of vehicles across areas in the network (congestion), the average wait or detour times of travellers, fleet utilization, or the level of sharing in a pooling service. Due to the complexity of the TNC operations, non-linearity of some relationships and heterogeneity of the client population are not always accounted in models. As a result, mathematical models, are not always capable of capturing the highly complex dynamics of the TNC environment and are subject to further validation.
 
 To complement the scrutiny of their algorithms and capture higher-order effects, modellers resort to digital twins or simulations, using Agent-Based Models (ABMs). Through the use of ABMs, researchers can apply their algorithms in a realistic environment, observe and quantify the emergent behaviour of the system in time. ABMs are bottom-up models, which simulate the behaviour of actors (clients, vehicles, operators) in the system by defining them as software objects (agents), which interact according to behaviours set by the modeller. As a result, they are dis-aggregated representations of reality, which allow for heterogeneous characteristics between agents, and are capable of simulating highly complex and non-linear systems \citep{doi:10.1057/jos.2010.3}.
 
 An example of an AV TNC problem which requires the use of ABMs is critical fleet sizing. Critical fleet sizing refers to the minimum fleet size required for the rate of incoming requests to be sustainably serviced. Specifically, suppose incoming requests join a queue until an available vehicle is assigned to them. In that case, fleet sizes below the critical, contribute to increasing queue lengths with incoming requests, therefore resulting in spiking assignment wait times for customers. This problem has been traditionally tackled in research through the use of queuing theoretical models with multiple servers by using parameters such as the rate of request arrivals and the average service time. However, the underlying network structure, the variation of velocities throughout the network and the spatial distribution of origin-destination locations contribute to non-homogeneous service times which make aggregation a challenging task. Furthermore, as the pickup time after assignment varies with fleet size, it results in an additional non-linear component in the service time. As a consequence, the more reliable route for estimating critical fleet sizes in TNC systems is the use of simulations which capture the complexities mentioned above.
 
 In reviewing the relevant literature in Section \ref{sec: Literature}, we observe that researchers applied ABMs extensively to investigate the impacts of AV TNC operations. Nonetheless, in anticipation of time-consuming software development, many studies resort to the use of ABM software packages which are inflexible when it comes to modelling non-trivial operational strategies. By contrast, studies which apply bespoke ABMs to test sophisticated optimization techniques often lack clarity in the structure of their simulation, which hinders reproducibility of their research and could potentially lead to significant omissions in the implementation. We therefore find that the existing state-of-the-art on AV TNC simulations, lacks a modular framework describing the components and building sequence of such ABMs for the different problems entangled in autonomous ride-sourcing, which is required to streamline such research.
 
 To address this literature gap, we develop an agent-based modelling framework which identifies the core components of ABMs in autonomous ride-sourcing systems. We model our framework using the fundamentals of agent-based modelling and propose a building sequence for structuring simulations of the various problems in autonomous ride-sourcing. To demonstrate the practicality of our ABM framework, we use it to investigate the critical fleet size problem and compare the results with an aggregated model based on the theoretical properties of stable queues with multiple servers. Our contribution is summarized as follows:
 \begin{itemize}
     \item We anatomize the core components of ABMs in autonomous ride-sourcing systems and present them in clear-cut classifications based on the fundamentals of agent-based modelling.
     \item We provide a simulation building sequence for tackling the different problems identified in autonomous ride-sourcing.
     \item We propose a model which makes use of pickup wait times in identifying the minimum required fleet size for autonomous ride-sourcing fleets based on queuing theoretical implementations. 
     \item We perform simulations with an ABM, which we structure using our proposed framework, to identify upper bounds of the critical fleet size for the urban areas of Manhattan, San Francisco, Paris and Barcelona. We then compare the results with the expected outputs of our aggregated models.
 \end{itemize}
 
The remainder of this paper is structured as follows: in Section \ref{sec: Literature}, we provide a literature review of the ABM implementations for autonomous ride-sourcing systems. In Section \ref{sec: ABM}, we outline the principles of agent-based modelling and provide examples of the approach across different disciplines. We present our ABM Framework and the model building sequence for different autonomous ride-sourcing problems in Section \ref{sec: Framework}. In Section \ref{sec: Queue} we provide an outline of the literature on fleet sizing and present queuing theoretical formulations to highlight the areas of additional complexity with a pickup wait time model. We perform a series of simulations to identify critical fleet sizes and maximum pickup wait times using our ABM framework and discuss the results in Section \ref{sec: Discussion}. Conclusions and recommendations for future work are provided in Section \ref{sec: Conclusion}.

\section{Related Work} \label{sec: Literature}

In this section we present the breadth of ABM applications in AV ride-sourcing and fleet sizing. To do so, we split the areas of application into the assignment problems of vehicle dispatching and ride-sharing, electric vehicle charging, fleet sizing, mode choice and pricing, and idle vehicle rebalancing. Table \ref{tab:Literature} summarizes the categorization of related literature on ABM applications for autonomous ride-sourcing.

Considering the dispatching problem, \cite{shen2015managing} used an ABM to evaluate different dispatching strategies to improve wait times and trip success rate of private trips. In terms of ride-sharing only, \cite{levin2017general} used dynamic ride-sharing heuristics and a cell transmission model in an ABM to simulate the effects of ride-sharing on traffic flow. \cite{dia2017autonomous} applied an ABM to compare the impact between trip sharing in an Autonomous Mobility on Demand (AMoD) scheme to private vehicle ownership.

The concept of Shared Autonomous Vehicles (SAVs), an autonomous extension of the car-sharing industry, which resembles AV TNCs, was introduced by \cite{fagnant2013travel}. The authors used a bespoke ABM to evaluate the impact of empty vehicle rebalancing in an SAV fleet in Austin, Texas. Their work was extended in a study by \cite{fagnant2015operations}, where they applied MATSim, an open-source ABM traffic simulation software, to evaluate the effects of different rebalancing strategies on customer wait times and fleet costs for SAV fleets. \cite{boesch2016autonomous} also used MATSim to identify optimal fleet sizes for different levels of demand for an SAV fleet. 

\begin{table}[h]
\centering
\scriptsize
\caption{Summary of related literature on ABM applications for autonomous ride-sourcing.}
\label{tab:Literature}
\begin{tabular}{@{}lccccc@{}}
\toprule
{\color[HTML]{000000} ABM   Studies} &
  {\color[HTML]{000000} \begin{tabular}[c]{@{}l@{}}Dispatching, \\ Sharing\end{tabular}} &
  {\color[HTML]{000000} Charging} &
  {\color[HTML]{000000} Fleet Size} &
  {\color[HTML]{000000} \begin{tabular}[c]{@{}l@{}}Mode Choice, \\ Pricing\end{tabular}} &
  {\color[HTML]{000000} Rebalancing} \\ \midrule
{\color[HTML]{000000} \begin{tabular}[c]{@{}l@{}}\cite{fagnant2013travel}   \\ \cite{fagnant2015operations} \\ \cite{zhang2016control} \\ \cite{wen2017rebalancing} \\ \cite{alonso2017predictive} \\ \cite{lin2018efficient} \\ \cite{wallar2018vehicle} \\ \cite{iglesias2018data}\end{tabular}} &
  {\color[HTML]{000000} } &
  {\color[HTML]{000000} } &
  {\color[HTML]{000000} } &
  {\color[HTML]{000000} } &
  {\color[HTML]{000000} \checkmark} \\ \midrule
{\color[HTML]{000000} \cite{boesch2016autonomous}} &
  {\color[HTML]{000000} } &
  {\color[HTML]{000000} } &
  {\color[HTML]{000000} \checkmark} &
  {\color[HTML]{000000} } &
  {\color[HTML]{000000} } \\ \midrule
{\color[HTML]{000000} \begin{tabular}[c]{@{}l@{}}\cite{shen2015managing}   \\ \cite{levin2017general} \\ \cite{dia2017autonomous}\end{tabular}} &
  {\color[HTML]{000000} \checkmark} &
  {\color[HTML]{000000} } &
  {\color[HTML]{000000} } &
  {\color[HTML]{000000} } &
  {\color[HTML]{000000} } \\ \midrule
{\color[HTML]{000000} \begin{tabular}[c]{@{}l@{}}\cite{martinez2015urban}   \\ \cite{martinez2016shared} \\ \cite{horl2017agent}\end{tabular}} &
  {\color[HTML]{000000} \checkmark} &
  {\color[HTML]{000000} } &
  {\color[HTML]{000000} } &
  {\color[HTML]{000000} \checkmark} &
  {\color[HTML]{000000} } \\ \midrule
{\color[HTML]{000000} \begin{tabular}[c]{@{}l@{}}\cite{chen2016management}   \\ \cite{maciejewski2016congestion}\end{tabular}} &
  {\color[HTML]{000000} } &
  {\color[HTML]{000000} } &
  {\color[HTML]{000000} } &
  {\color[HTML]{000000} \checkmark} &
  {\color[HTML]{000000} } \\ \midrule
{\color[HTML]{000000} \cite{chen2016operations}} &
  {\color[HTML]{000000} } &
  {\color[HTML]{000000} \checkmark} &
  {\color[HTML]{000000} } &
  {\color[HTML]{000000} } &
  {\color[HTML]{000000} \checkmark} \\ \midrule
{\color[HTML]{000000} \cite{maciejewski2016large}} &
  {\color[HTML]{000000} \checkmark} &
  {\color[HTML]{000000} } &
  {\color[HTML]{000000} \checkmark} &
  {\color[HTML]{000000} } &
  {\color[HTML]{000000} } \\ \midrule
{\color[HTML]{000000} \cite{liu2017tracking}} &
  {\color[HTML]{000000} } &
  {\color[HTML]{000000} } &
  {\color[HTML]{000000} \checkmark} &
  {\color[HTML]{000000} \checkmark} &
  {\color[HTML]{000000} } \\ \midrule
{\color[HTML]{000000} \cite{fagnant2018dynamic}} &
  {\color[HTML]{000000} \checkmark} &
  {\color[HTML]{000000} } &
  {\color[HTML]{000000} \checkmark} &
  {\color[HTML]{000000} } &
  {\color[HTML]{000000} \checkmark} \\ \bottomrule
\end{tabular}
\end{table}

The impact of vehicle pooling strategies on mode choices was investigated by \citep{martinez2015urban}, \cite{martinez2016shared} and \cite{horl2017agent} using ABMs. Other studies, such as \citep{chen2016management} and \citep{maciejewski2016congestion}, attempted to identify the SAV mode penetration in the market using different pricing schemes. \cite{chen2016management} used a bespoke ABM whereas \cite{maciejewski2016congestion} used MATSim. The work in \cite{chen2016operations} extended the study in \cite{chen2016management}, to investigate the operational costs for Shared Autonomous Electric Vehicle (SAEV) fleets and vehicle rebalancing. 

The problem of identifying a cost-efficient fleet size was investigated in combination with other TNC/SAV in studies such as \citep{maciejewski2016large}, \citep{liu2017tracking} and \citep{fagnant2018dynamic} using ABMs. \cite{maciejewski2016large} employed a large-scale micro-simulation of SAVs in Berlin and Barcelona via MATSim, to evaluate the effectiveness of dispatching strategies on the required fleet size to serve the trip data-set used. \cite{liu2017tracking} also used MATSim to identify the effectiveness of pricing strategies on mode choice and fleet size in an SAEV fleet. \cite{fagnant2018dynamic} used an ABM to evaluate the impact of the level of ride-sharing and vehicle redistribution strategies on the required fleet size for an SAV operation.

Studies focused on sophisticated vehicle rebalancing strategies such as queuing theoretical models and aggregated optimization methodologies resorted to bespoke simulation models for validation rather than open-source ABM software such as MATsim. \cite{zhang2016control} used an ABM to validate their queuing theoretical vehicle redistribution model whereas \cite{wen2017rebalancing} and \cite{lin2018efficient} validated a reinforcement learning redistribution approach using an ABM. Similarly, \cite{alonso2017predictive}, \cite{wallar2018vehicle} and \cite{iglesias2018data} validated their aggregated optimization approach for vehicle redistribution using an ABM.  

\section{Agent-based Modelling Principles} \label{sec: ABM}
The seminal work for agent-based modelling could be attributed to John von Neumann and Stanislaw Ulam when they created a grid-based model as a means for the design of the universal constructor; a machine capable of self-replication. Their grid-based model was comprised of a limited number of cells, each in one of a finite collection of states. The cells were assumed to alternate states in an iterative procedure, following a fixed rule. This type of discrete grid-based model was later termed cellular automata (CA) \citep{von1966theory}. The concept of CA was subsequently incorporated in John Conway's Game of Life which is also one of the earliest applications of agent-based modelling \citep{berlekamp2018winning}.

What both Von Neumann's and Conway's experiments were showcasing, was that complex macro patterns could emerge by defining microscopic interaction rules between simple individual entities. These fundamental principles of CAs, extend to describe agent-based modelling, which, according to \cite{bonabeau2002agent}, is the practice of describing a system by defining a collection of self-determining agents. Unlike Conway's Game of Life; however, the agents are not necessarily restricted to simplistic behaviour, since the level of detail in the underlying system is entirely dependent upon the modeller. As such, agent-based models constitute powerful tools for modelling highly complex systems to identify emergent phenomena by enabling heterogeneity in the agent population and non-trivial networks of communication.

\subsection{Key Agent-Based Modelling Components}
According to \cite{doi:10.1057/jos.2010.3}, agent-based models have three key components:
\begin{itemize}
    \item A set of agents.
    \item The network of connectedness between agents.
    \item The model's environment.
\end{itemize}

The agents are defined by their individual properties and their collective set of states. Behavioural definitions govern the alterations between states over time. Agent behaviour can range from fixed and simplistic (adhering to a finite set of rules), to adaptive and intelligent (perceptive with utility maximization). \cite{doi:10.1057/jos.2010.3} also argue that agents need to be uniquely identifiable and modular. They explain modularity as the property which clearly defines a boundary on what constitutes an agent in a model.

The network of connectedness between agents is otherwise referred to as the topology of the model \citep{bonabeau2002agent, doi:10.1057/jos.2010.3}. It represents the mapping of possible social interactions. Each agent interacts with its neighbours, typically a subset of other agents in the model. The extent of neighbours (accessible agents), can vary according to the nature of the underlying topological model. In CA models such as the Game of Life, the topology is grid-based, with agents only connected to their immediate eight neighbouring cells. In models which consider a Euclidean topology, agents may only interact with others in their vicinity, up to a maximum Euclidean distanced radius. If the modeller considers a graph topology, agents (represented as nodes) are connected to their neighbours by graph edges.

\begin{figure}[h]
\centering
\includegraphics[width=0.6\textwidth]{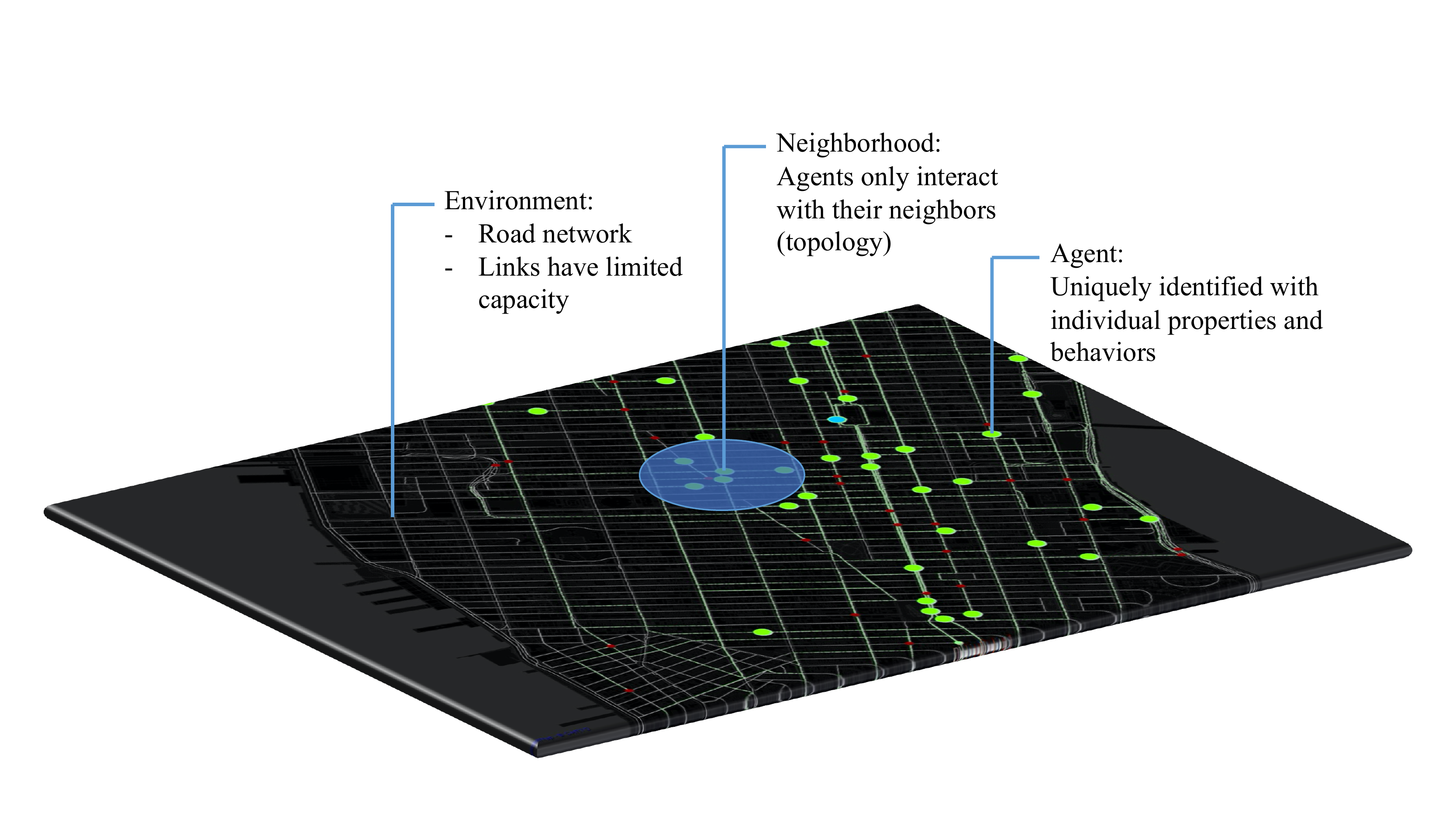}
\caption{Illustration of agent-based model components in a road network environment.}
\label{fig:2.1}
\end{figure}

The environment is used in order to represent the spatial and temporal dimensions of the model. It is, therefore, the boundary of the existence of everything within an agent-based model. As such, agents' locations in time are derived by the environment. Also, complex environments could be characterized by local properties, such as events or limited resources, which create restrictions and impact agents \citep{crooks2012introduction}. For example, in an agent-based model which simulates motor traffic, the environment would take the form of a road network. Consequently, road links would be subject to a limited capacity of vehicles, and an event such as a storm, might also reduce the utility of travel. Figure \ref{fig:2.1} illustrates the agent-based model components described in a road network environment.

\subsection{Agent-Based Modelling Across Disciplines}
The modularity and generality of the agent-based approach make it suitable for implementation in a variety of disciplines. Additionally, the substantial technological advancements in computing and the growing amount of data availability in the last decade has facilitated the emergence of numerous agent-based modelling projects, notably in areas such as transport, economics and epidemiology. In the following sub-sections, we offer an overview of some examples of agent-based methodologies applied in these areas.

\subsubsection{Transport}
Transportation planners are typically interested in evaluating the impact of new implementations, such as policy or infrastructure projects on transportation networks. These implementations could range from the introduction of new streets, congestion pricing or even public transit lines in existing networks. Traffic engineers traditionally derive key performance indicators of impact such as traffic flows, environmental footprint and cost-benefit ratios using aggregated four-step forecasting models\footnote{Four stage models include trip generation, trip distribution, modal split and traffic assignment \citep{mcnally2000four}.}. Nonetheless, the emergence of intelligent transportation systems, as well as complex solutions such as intersection design, created the need for modelling tools which assess the impact of such systems from a dis-aggregated perspective \citep{Auld2016}. 

Agent-based simulations have therefore been useful in complementing the traditional four-step models due to their decentralized nature. Software packages such as Vissim \citep{fellendorf2010microscopic} allow micro-simulation of road networks where vehicles and pedestrians can act as agents in a realistic three-dimensional environment. The three-dimensional component of Vissim, however, prohibits the modelling of large scale networks. Nevertheless, software such as TRANSIMS \citep{nagel1999transims} and MATSim \citep{balmer2006agent} can conduct agent-based simulations of city-scale models, by considering vehicles and travellers as agents, but omitting the low level three dimensional network detail of Vissim.

\subsubsection{Economics}
In macro-economics, traditional analysis utilised the dynamic stochastic general equilibrium (DSGE) model. In these models, economists aim to identify the effects of policy on growth and business cycles in economies. Their premise is to evaluate the evolution of econometric factors through time, assuming that the economy is subject to random shocks and also in Walrasian equilibrium \citep{burgess2013bank, fagiolo2019validation}. Such models have been criticised by various studies \citep{mankiw2006macroeconomist, haldane2016dappled, fagiolo2017, fagiolo2019validation} for invalid assumptions additional to equilibrium, such as population homogeneity and rationality\footnote{DSGE models typically assume a representative agent who is entirely rational.}. 

The simplicity of the DSGE assumptions leads to their inability to describe highly non-linear effects which are results of emergent behaviour or wealth distribution in an economy. As a consequence, agent-based modelling in finance is a useful tool to explore systems which are not in equilibrium and involve heterogeneous agents which act based on heuristics\footnote{In behavioural economics it has been argued that people use heuristics in decision making rather than exact optimisation\citep{simon1955behavioral}.}\citep{turrell2016agent}. Some implementations of agent-based models in finance consider the non-linear relationships of corporate bond trading in mutual funds \citep{braun2016agent} and the underlying agent interactions that govern the UK housing market\citep{baptista2016macroprudential}.

\subsubsection{Epidemiology}
During the unfolding of the Covid-19 pandemic, especially in the early months of 2020, agent-based modelling was critical in policy implementation in the United Kingdom and the United States \citep{adam2020special}. Agent-based models built in cooperation between engineers, mathematicians and epidemiologists, assisted in determining the footprint of the SARS-CoV-2 virus on the population. Studies such as the one by \cite{ferguson2020report} influenced the UK government in implementing strict social distancing rules to prevent an anticipated sharp increase in hospitalization rates which would overwhelm the National Healthcare System (NHS). The data input of such epidemiological agent-based models has been data-heavy but also uncertain. Consequently, stochastic processes were incorporated in the models, and multiple runs of the simulations assured the consideration of numerous scenarios by the modellers. 

The brief structure of models, such as the ones employed to inform governments, was comprised of agents clustered in individual households. Socio-economical characteristics from census studies were employed to characterize the properties of the agent population. Some of the most critical properties have been age, sex, health status, type of work and amount of contacts. Then, the modellers anticipated social mixing using data such as modes of travel and origin-destination matrices. The topology of these models was defined by agent geographic vicinity, as the transmission of the virus is only possible with the proximity of agents in the underlying environment. Additional data inputs, considered the biological description of the virus, such as the transmission rate and the viral impact on each agent \citep{ferguson2020report, chang2020modelling}.

\section{Agent-Based Modelling Framework} \label{sec: Framework}
We structure our ABM framework to facilitate a streamlined approach in creating models for the relevant problems in autonomous ride-sourcing highlighted in Section \ref{sec: Literature}. As such, we first identify and analyse all the core components of the framework in line with the fundamental description provided in Section \ref{sec: ABM}, and then we present our model building sequence.

\subsection{Framework Components} \label{frame: components}
Following the terminology in the work of \cite{doi:10.1057/jos.2010.3}, we split our framework composition between the components of agents, their respective topologies and the environment in which they exist. 
\subsubsection{Types of Agents} \label{frame: agents}
We used the research outlined in Section \ref{sec: Literature} to derive the core agents in an AV TNC system. In doing so, we identified three main types of agents; the traveller, the vehicle and the operator. We assume that each type of agent is characterised by some distinct properties and behaviours. However, to ensure uniqueness and modularity, all agents across all types have a unique identifier property and a list of states. We term the list of states as well as the mechanism description for alternating between states as state logic.

Furthermore, we identify a list of KPIs for each agent which could serve as a model output for each agent. Table \ref{tab: agents} presents these core agents, their properties, behaviours and KPIs. Properties, behaviours and KPIs in our framework are non-exhaustive and also non-binding for model building but represent the typical ones encountered in the literature.

\begin{table}[]
\centering
\caption{Description of core agents' properties, behaviours and key performance indicators (KPIs).}
\label{tab: agents}
\begin{tabular}{@{}llll@{}}
\toprule
Agent &
  Properties &
  Behaviours &
  KPIs \\ \midrule
Traveller &
  \begin{tabular}[c]{@{}l@{}}- Unique identifier\\ - Origin and destination\\ - Request time\\ - Utility function\\ - Reservation price\end{tabular} &
  \begin{tabular}[c]{@{}l@{}}- Mode choice\\ - State logic\end{tabular} &
  \begin{tabular}[c]{@{}l@{}}- Utility\\ - Wait time\\ - Travel time\\ - Detour\\ - Cost\\ - Trips shared\end{tabular} \\ \midrule
Vehicle &
  \begin{tabular}[c]{@{}l@{}}- Unique identifier\\ - Location\\ - Capacity\\ - Velocity\\ - Range and charge\\ - Schedule\\ - Revenue function\\ - Cost function\end{tabular} &
  \begin{tabular}[c]{@{}l@{}}- Charging station choice\\ - Parking station choice\\ - State logic\end{tabular} &
  \begin{tabular}[c]{@{}l@{}}- Mileage\\ - Revenue\\ - Cost\\ - Trips served\\ - Service rate\end{tabular} \\ \midrule
Operator &
  \begin{tabular}[c]{@{}l@{}}- Unique identifier\\ - Fleet (vehicles)\\ - Assignment strategies\\ - Pricing strategies \\ - Routing algorithm\end{tabular} &
  \begin{tabular}[c]{@{}l@{}}- Assignment/pricing choice\\ - Fleet management strategy\\ - State logic\end{tabular} &
  \begin{tabular}[c]{@{}l@{}}- Mileage\\ - Revenue\\ - Cost\\ - Trips served\\ - Service rate\end{tabular} \\ \bottomrule
\end{tabular}
\end{table}

Travellers are potential clients who submit requests to the TNC for rides. Consequently, throughout the ride-sourcing literature, the terms of clients, requests or riders have been interchangeable \citep{Wang2019}. Their goal is to travel from an origin location to a destination. As such, their origin and destination coordinates are static\footnote{Static properties do not change in the model.} properties. We assume travellers enter the system at their static request time and exit after they are delivered at their destination or when they abort the TNC service by cancelling their request (if this is possible in the model). Multiple travellers could also be bundled into the same request, with the number of persons in the request to be a property description.

To assist the acumen of traveller agents (if required), we assume a utility function that could take inputs which differentiate the quality of a TNC option amongst other options in the model (if any). Consequently, it could have deterministic inputs such as the prospective wait time, travel time and service price. The capability to support heterogeneous populations in ABMs, allows for monetising parameters in the utility function, such as the value of travel time, wait time, service price or other inputs, to be specific to individual travellers. Modellers could also define other traveller specific properties such as a reservation price, which reflects the maximum amount a traveller is willing to pay for a travel option. 

Vehicles represent the AVs in the system, and their goal is to serve travellers. They can have properties such as the total and current capacity, as well as the total and current range\footnote{Potential mileage based on electric charge level.}. Since vehicles are self-moving agents, they are also characterised by velocity and location properties, which depend on the environment and vary during the simulation. We also assume that operators own the vehicles due to their autonomous nature. Consequently, their activity schedule and revenue function are both governed by the assignment and pricing strategies of their operator. The type of vehicle also defines its cost function, which could take as input the mileage and velocity fluctuations during the simulation. 

Operators are agents which denote the AV fleet owners in the ABM. Depending on the model, there could be one or multiple operators in the system (monopolistic vs competitive scenarios). They dictate the assignment, pricing and routing methods which vehicles follow during the simulation. Operators are not explicitly regarded as agents in the literature outlined in Section \ref{sec: Literature}; however modelling them as software entities allows for extensible properties which enable the simulation of scenarios that emulate adaptive behaviour by TNCs. Such scenarios could be the response to a congestion charge by a regulator agent or the fluctuation of pricing in competition. 

As we mentioned earlier, all agents in our framework are assumed to have an arbitrary set of states. These states alternate during the simulation for each agent depending on discrete events, processes (i.e. assignment heuristic) or decisions. The complexity of decision making ca vary. The majority of ABMs cited in Section \ref{sec: Literature} assume travellers always request and follow an assignment from a TNC unless some thresholds such as wait time, detour time or a reservation price are exceeded. Other studies \citep{chen2016management, liu2017tracking} assume more sophisticated models, in which each option's utility is an input to a discrete choice model. Discrete choice models have not been widely applied on AV TNC simulation studies, mainly due to the lack of necessary population information to validate such models.

Vehicle behaviours are governed by operator properties, such as assignment and pricing strategies, as well as the routing algorithm. Typically, a vehicle might be preliminarily assigned to a traveller by the operator using some assignment heuristic. The traveller then would either accept or abort the TNC service. Upon acceptance, the vehicle will follow a route to the traveller origin based on the routing algorithm specified by the operator. In cases where a vehicle needs to recharge or park, vehicles could decide which charging or parking station to visit using simple heuristics such as the nearest available station. 

Our ABM framework does not intend to define the agents' behaviours but rather collectively classify them as sub-systems. We follow this approach to maintain a flexible and extensible nature in modelling. Nonetheless, in Table \ref{tab: States}, we provide some basic states for travellers and vehicles based on the models presented in Section \ref{sec: Literature}. Table \ref{tab: States} does not provide any example states for operators due to the lack of applications in research. We also note that we did not include any vehicle states related to electric charging, but these states could be implemented as extensions of the core state logic. 

\begin{table}[h]
\centering
\caption{Core state description for travellers and vehicles.}
\label{tab: States}
\begin{tabular}{@{}lll@{}}
\toprule
Agent                    & State                & Description                                                                                                 \\ \midrule
\multirow{5}{*}{Traveller} &
  Waiting for assignment &
  \begin{tabular}[c]{@{}l@{}}The traveller has requested a ride but has not\\ been assigned to a vehicle.\end{tabular} \\ \cmidrule(l){2-3} 
 &
  Waiting for pick-up &
  \begin{tabular}[c]{@{}l@{}}A vehicle has been assigned to the traveller and\\ the traveller waits to be picked up.\end{tabular} \\ \cmidrule(l){2-3} 
                         & In trip              & The traveller is in the vehicle.                                                                            \\ \cmidrule(l){2-3} 
                         & Served               & \begin{tabular}[c]{@{}l@{}}The traveller has been delivered to the \\ destination location.\end{tabular}    \\ \cmidrule(l){2-3} 
                         & Aborted              & \begin{tabular}[c]{@{}l@{}}The traveller aborted the service before pick-up.\end{tabular}                \\ \midrule
\multirow{5}{*}{Vehicle} & Idle                 & The vehicle does not have any tasks.                                                                        \\ \cmidrule(l){2-3} 
                         & Travelling to origin & \begin{tabular}[c]{@{}l@{}}The vehicle is travelling to a traveller's origin\\ location.\end{tabular}      \\ \cmidrule(l){2-3} 
                         & Loading              & \begin{tabular}[c]{@{}l@{}}The traveller is boarding the vehicle at the \\ origin location.\end{tabular}    \\ \cmidrule(l){2-3} 
 &
  Travelling to destination &
  \begin{tabular}[c]{@{}l@{}}The vehicle is travelling to the traveller's \\ destination location.\end{tabular} \\ \cmidrule(l){2-3} 
                         & Unloading            & \begin{tabular}[c]{@{}l@{}}The traveller is exiting the vehicle at the \\ destination location.\end{tabular} \\ \bottomrule
\end{tabular}
\end{table}

\subsubsection{Topologies} \label{frame: top}
We have previously defined topology in Section \ref{sec: ABM} as the infrastructure enabling interactions between agents. We identified possible topology settings which are based on euclidean distance or graph structures; however, in our AV TNC modelling framework, we propose multiple levels of topologies which are also more complex than their one-dimensional counterparts mentioned in Section \ref{sec: ABM}. 

To accentuate the underlying topologies in a TNC ABM, we first make notice of the interactions between types of agents in a realistic TNC service. The typical process involves a traveller submitting a request to an operator. The operator, in turn, assigns a vehicle and informs the traveller of the expected pick-up time. If the traveller accepts the assignment, the vehicle is then instructed by the operator to pick-up its assigned traveller. It is therefore apparent that the operator brokers any agent interaction. Travellers do not directly interact with nearby vehicles, nor do vehicles decide which travellers to select, but the operators instead instruct them.

It is also evident that different topologies are relevant at different times for different agents. On request submission, travellers can access all TNC operators. During the assignment, the operator structures a graph topology which outlines all possible assignments between travellers and vehicles (and also travellers and travellers in the case of ride-sharing). The assignment graph topology could depend on traveller requirements, vehicle properties, as well as the geographical vicinity between these agents. Furthermore, the aforementioned topology includes both travellers and vehicles, but it is only actionable by the operator. Upon preliminary assignment, each operator informs the travellers of their option, and then if travellers accept the offer, the operator instructs the vehicles to move to their assigned traveller origin locations.

The processes described above, highlight an open-access topology between travellers and operators, which is relevant at request submission, and upon traveller choice when offers are presented. We also define a preliminary topology between vehicles and travellers during the assignment. We term the assignment topology as preliminary, since it offers no access between vehicles and travellers, but is only visible to the operator. Derived from the preliminary topology, we also identify a one-to-one topology between travellers and their assigned vehicles after acceptance by the travellers.

Due to the technological nature of request submission to a TNC platform, we assume no direct interaction between travellers and nearby vehicles, unless the operator decides an assignment. Nonetheless, such a simplified geographical topology could be implemented if modellers aim to test competition between TNC platforms and street hailing taxis. We also assume no direct interaction between the same types of agents, such as traveller-traveller\footnote{Traveller-traveller connectedness could be apparent during a ride-sharing assignment process, but there is no communication between travellers.} or vehicle-vehicle\footnote{Vehicle-vehicle connectedness could be useful in the connected-autonomous vehicle (CAV) problem for traffic purposes but lies beyond the AV TNC scope.} connectedness.

The implementations of assignment strategies in the research outlined in Section \ref{sec: Literature} vary from simple First-In-First-Out (FIFO) heuristics to aggregate optimization assignments. In a FIFO heuristic, travellers enter the end of a queue of unassigned travellers upon request submission. Then, the operator assigns the closest vehicle to each traveller sequentially, from first to last in the queue. Aggregate optimization assignment procedures were used mainly in studies proposing idle vehicle redistribution methodologies \citep{alonso2017predictive, lin2018efficient, wallar2018vehicle, iglesias2018data}. Assignments using optimization, are decided in intervals in a quasi-online approach. During these intervals, travellers' requests and available vehicles accumulate a list, and a cost minimization optimization program decides the optimal assignment at the end of the interval.  

To reduce the instance size of optimization problems for assignment, \cite{santi2014quantifying} proposed the use of shareability networks. Although the initial application was presented for use in identifying potentially shareable trips in a ride-sharing scheme, the concept can be extended to an assignment between travellers and vehicles for both shared and private trips. In such networks, nodes could represent traveller or vehicle agents, and edges between them could represent a capability for assignment subject to constraints. More sophisticated assignment strategies also incorporate pricing into the creation of such networks by using auctions \citep{james2018double, karamanis2020assignment}. The shareability networks or more simplified techniques, such as FIFO, constitute the preliminary topology visible to the operator. 

\subsubsection{Environment} \label{frame: env}
We set the road-network as the core environment in our ABM framework. The presence of the operator is abstract (not physical); nonetheless, travellers and vehicles physically exist in a road-network via origin/destination and current location, respectively. Some simulations might ignore the urban road-network and instead use grid-like environments representing physical locations or even euclidean space. However, using these alternative environments might be an over-simplification as vehicle properties such as velocity and schedule are derived from road-network information. Furthermore, processes such as assignment, pricing and routing utilize road-network information via shortest path algorithms.

The main road-network components are road-nodes and road-links. We assume road-nodes have the fundamental property of coordinates, and edges have the fundamental properties of an inbound and outbound road-node, as well as a geometry indicated by a list of coordinates. Information such as road-node neighbours, or the length of each road-link could be derived from the fundamental properties above. Additional to these properties (but not necessarily essential), could be a specific velocity or a velocity vs traffic density profile in each road-link. Simplified models could assume an average velocity over an entire road-network instead. Background traffic could also be implemented as a randomized input in a velocity-traffic density profile for realistic ABM implementations. 

By assuming a velocity-traffic density profile for each edge, a modeller can define space in each link as a limited resource. Furthermore, additional environmental features on top of the core road-network structure could be capacitated parking or charging stations, situated in selected road-nodes across the road-network. Such stations also constitute a time-varying limited resource in an ABM. Additional environmental features which enhance complexity could include networks for different modes (public transport, cyclic, walking) or external events which impact the properties of the environment (i.e. extreme weather).

\subsection{Model Building} \label{sec: Build}
The components introduced in Section \ref{frame: components} serve as the raw materials required in ABM building for autonomous ride-sourcing problems. The amount of necessary detail in agents, topologies and environment composition could vary significantly depending on the ABM scope. Therefore, deciding the model contents as well as the sequence in which the building blocks should be created is essential in avoiding computational burden and over-design of the system.

We first highlight that the various versions of agents, as well as the environment's building blocks, can contain information in the form of properties. Examples of properties for agents include the origin and destination locations of travellers and the locations and capacities of vehicles. Furthermore, environment components such as road-nodes and charging stations have the property of coordinates. Agents also utilize behavioural processes, such as mode choice, charging/parking station choice, routing and assignment. Consequently, the preceding descriptions of agents and environmental components conveniently fit the description of software objects with Object-Oriented Programming (OOP) principles. 

In OOP, objects are software entities (classes) which hold properties and processes (methods). In that respect, we regard OOP languages as ideal for building ABMs. We classify objects into agents and non-agents since only agents have behavioural properties. Furthermore, we also categorize processes into decision problems and functions with scalar outputs. Functions with scalar outputs define scalar agent properties described in Table \ref{tab: agents}, such as traveller utility, vehicle revenue and cost, and serve as prerequisites to decision problems such as mode choice, routing and assignment. Table \ref{tab: objects and processes} summarizes the categorization of objects and processes in line with the descriptions presented in Section \ref{frame: components}.

\begin{table}[]
\centering
\caption{Categorization of objects and processes.}
\label{tab: objects and processes}
\begin{tabular}{@{}llll@{}}
\toprule
\multicolumn{2}{c}{Objects} &
  \multicolumn{2}{c}{Processes} \\ \midrule
Agents &
  Non-Agents &
  Decision Problems &
  Scalar Functions \\ \midrule
\begin{tabular}[c]{@{}l@{}}- Travellers\\ - Vehicles\\ - Operators\end{tabular} &
  \begin{tabular}[c]{@{}l@{}}- Road-network\\ - Road-node\\ - Road-link\\ - Infrastructure\end{tabular} &
  \begin{tabular}[c]{@{}l@{}}- Mode choice\\ - Routing\\ - Assignment\\ - Charging/Parking choice\\ - Fleet management\end{tabular} &
  \begin{tabular}[c]{@{}l@{}}- Utility\\ - Revenue\\ - Cost\end{tabular} \\ \bottomrule
\end{tabular}
\end{table}

\begin{figure}[h]
\centering
\includegraphics[width=0.6\textwidth]{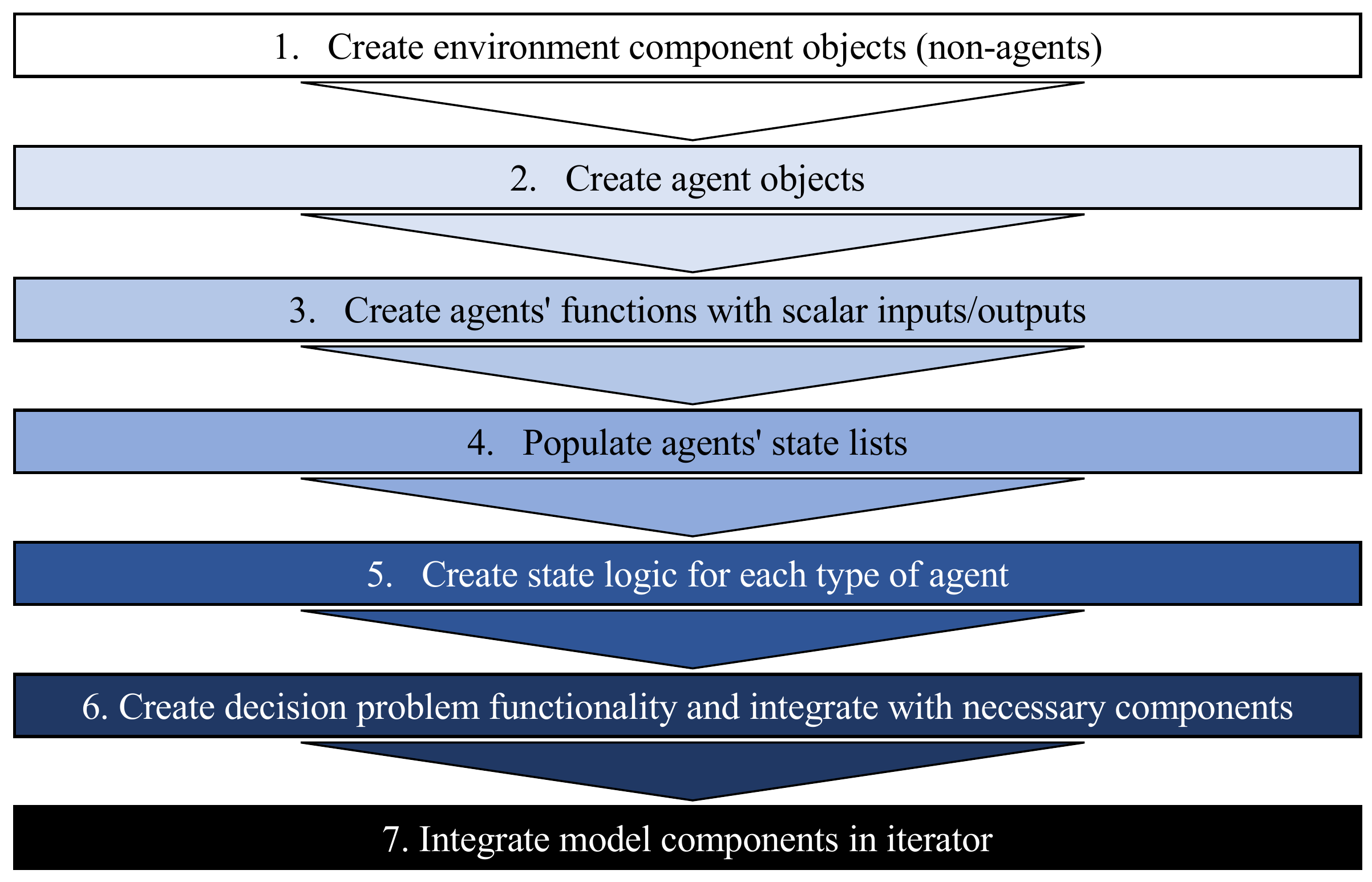}
\caption{Proposed model building sequence.}
\label{fig:4.2a}
\end{figure}

The categories proposed above also reflect the objects' complexity and processes, as well as our proposed sequence for model building. Decision problems present higher complexity than functions outputting scalar variables; agents are more complex than non-agents due to their respective behavioural and adaptive properties. Also, as stated before, decision problems need the outputs of scalar functions as inputs, and agents have properties derived from non-agent objects (environment components). As a consequence, our model building sequence proceeds from simple to complex software entities. We, therefore, propose the following sequence for model building, as shown in Figure \ref{fig:4.2a}.

% \begin{enumerate}
%     \item Create environment component objects (non-agents).
%     \item Create agent objects.
%     \item Create agents' functions with scalar inputs/outputs.
%     \item Populate agents' state lists.
%     \item Create state logic for each type of agent.
%     \item Create decision problem functionality and integrate with necessary components.
%     \item Integrate model components in iterator.
% \end{enumerate}

The sequence of operations during a simulation run is also non-trivial. In practice, objects which exist throughout the simulation time should be instantiated before the iterator starts. As such, infrastructure objects (non-agents) should be instantiated first, followed by operators, and their vehicles in locations across the network (derived from infrastructure objects). Travellers could then be created during the iterations at their request times. Once new travellers are added to the system, a state logic step should be performed for each agent at each iteration step until all iterations are completed. 

We assume different models might require a different state logic step sequence when considering all types of agents. For example, at the beginning of an iteration step, a traveller requests a trip from an operator. The operator, in turn, might trigger the assignment process in the same iteration step. If at least a vehicle is available and the assignment process is not computationally expensive, it is reasonable to assume that the assignment would be instantaneous. In that case, the operator could report the assignment to the traveller, with the traveller choosing the preferred mode in the same iteration step as the request. 

However, in cases where the assignment is indeed expensive computationally, such as in specific ride-sharing settings \citep{karamanis2020assignment}, the events of triggering the assignment process and obtaining a result do not occur at the same time even in realistic situations. During this period, from triggering a computationally expensive process to obtaining a result in a ride-sourcing environment, other operations might be underway. For example, vehicles not involved in that process might be travelling to a destination, and travellers involved in a similar assignment process do experience wait time which could be essential in their final choice. As a consequence, the use of multi-threading programming to run such computationally expensive processes in parallel with the main iteration might be required.

\begin{figure}[h]
\centering
\includegraphics[width=1\textwidth]{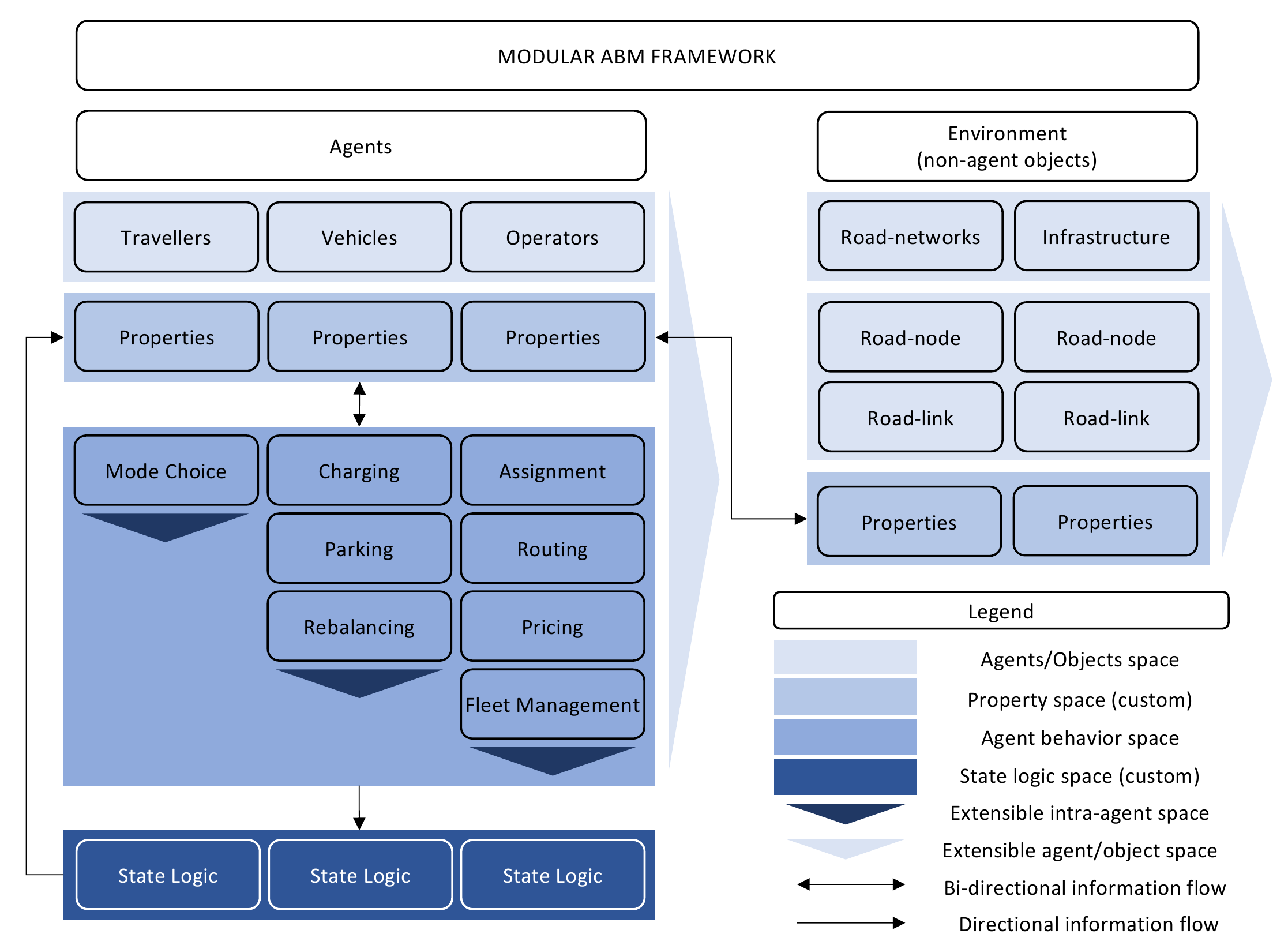}
\caption{Graphical illustration of the modularity of the proposed ABM framework.}
\label{fig:4.2b}
\end{figure}

\begin{figure}[h]
\centering
\includegraphics[width=0.7\textwidth]{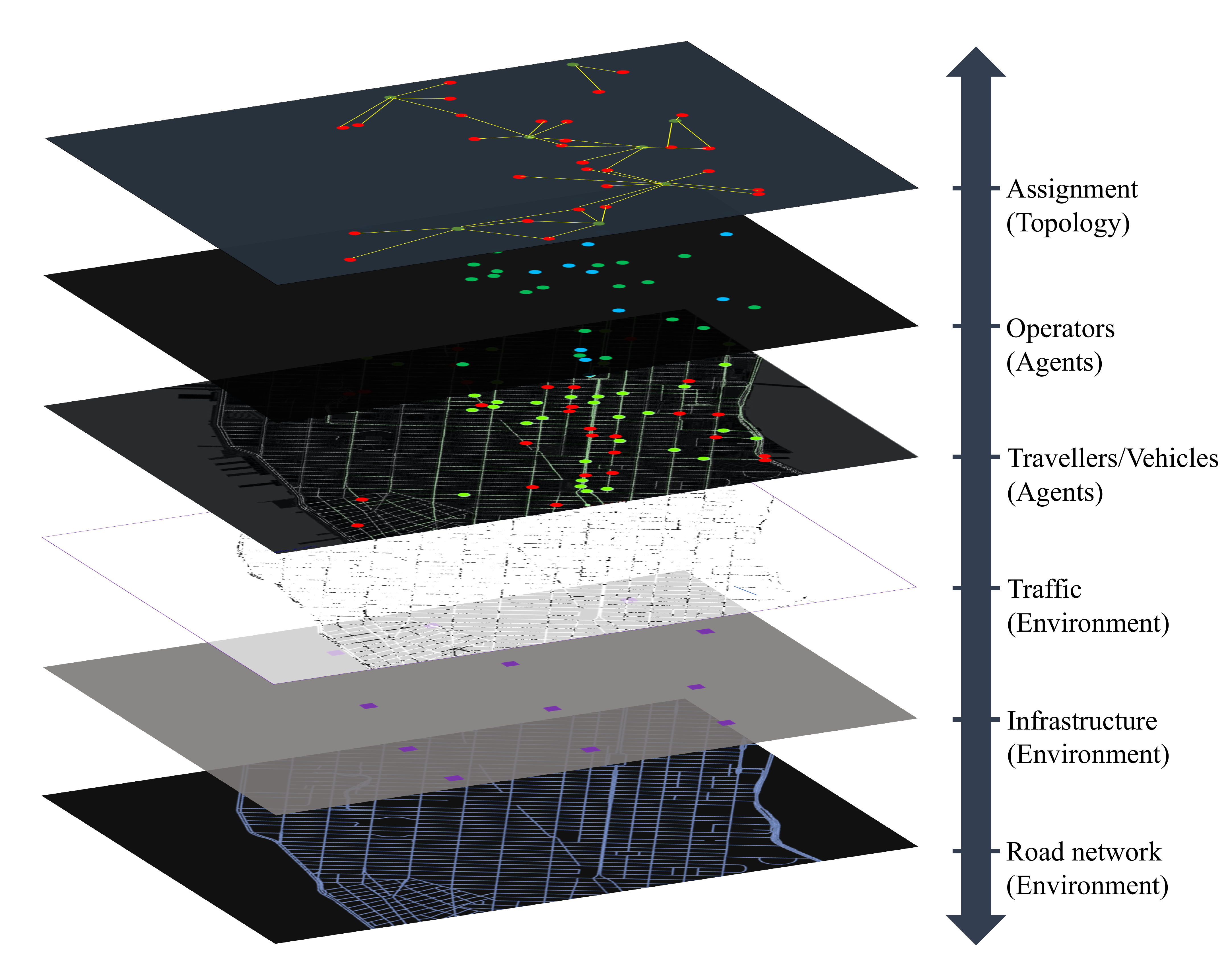}
\caption{Illustration of the extensible simulation layers in our ABM framework.}
\label{fig:4.1}
\end{figure}

The proposed ABM framework offers the capability of creating simulations with complexity according to the appetite of the modeller, by following a modular approach. Figure \ref{fig:4.2b} outlines the components of the proposed framework and its modular/extensible nature. Figure \ref{fig:4.1} outlines how the different layers of complexity blend together to create a realistic ABM for ride-sourcing. 

\section{Critical Fleet Size Modelling} \label{sec: Queue}
Various studies used queuing theory to model the operation of ride-sourcing services and successfully focused on identifying optimal idle vehicle redistribution strategies across networks \citep{zhang2016control, iglesias2019bcmp, braverman2019empty}. Such studies, model the arrival of travellers as a Poisson process in time, in which the travellers join a queue upon arrival and exit the system after they are served by vehicles, with service times usually modelled by an Exponential distribution. \cite{alonso2017predictive} and \cite{wallar2018vehicle} also focused on solving the empty vehicle redistribution problem, but instead applied aggregated optimization methods in combination with demand prediction to solve the problem. 

A key index of such vehicle redistribution studies is the average wait time which travellers experience until pickup, which is traditionally regarded as a quality of service metric in demand-responsive transport \citep{Yang2002}. We split traveller wait time in two categories, specifically, the waiting time from request to assignment, and the waiting from assignment to pickup. As such, assignment wait is governed by the number of vacant vehicles whereas waiting for pick up relates to the vicinity of those vacant vehicles. Consequently, understanding how the components of wait time vary with demand, vehicle availability and the spatio-temporal properties of the underlying network can be a useful input in fleet management models.

Our model assumes that travellers request trips through a central ride-sourcing fleet operator at a quasi-constant rate. Requests are added to a single queue upon arrival and are assigned to the closest available autonomous vehicle in the fleet using a first-in-first-out (FIFO) setting. For clarity of the demonstration of our ABM framework and identifying the properties of wait time variation, we do not consider acumen in traveller behaviour. Consequently, travellers exit the system once they are served, even if they experience significant wait times. We note that although this setting is unrealistic, in this case, is useful in identifying the fleet size limitations.

We assume an $M/M/c$ queue model as a way to represent the operation of the ride-sourcing fleet operator. The first and second $M$'s in the notation stand for Poisson generated traveller arrivals and exponentially distributed service times, respectively. Furthermore, $c$ denotes the number of parallel homogeneous servers (fleet size). As such, traveller arrivals occur at a rate of $\lambda$ per unit time and the average service time is set at $\Bar{t}$ units of time. Consequently, the service rate $\mu$ is equal to $1/\Bar{t}$. We, therefore, define the utilization rate $\rho$ of the fleet as follows:

\begin{equation} \label{eq:4.1}
    \rho = \frac{\lambda}{c\mu}
\end{equation}

As observed in \eqref{eq:4.1}, $c\mu$ denotes the number of travellers the fleet can serve in a period. As such, for $\rho > 1$, the traveller arrivals surpass the fleet's capacity, and the queue becomes unstable, with the average wait time in the queue $W_q$ growing to infinity. Consequently, for $\rho=1$ $c$ is equal to the minimum fleet size $c_0$ which can sustain the queue, given $\lambda$ and $\mu$. We, therefore, have that:

\begin{equation} \label{eq:4.2}
    c_0 = \frac{\lambda}{\mu}
\end{equation}

We define the mean number of travellers in the queue as $L_q$ (length of queue); similarly, we define the mean number of travellers in the system (including travellers being served) as $L$ and the average time in the system as $W$. Using Little's law \citep{ross2006introduction}, we have the following:

\begin{equation} \label{eq:4.3}
    L = \lambda W
\end{equation}

\begin{equation} \label{eq:4.4}
    L_q = \lambda W_q
\end{equation}

We also expect the following relationship between wait time in the queue the total time of travellers in the system:

\begin{equation} \label{eq:4.5}
    W = W_q + \frac{1}{\mu}
\end{equation}

For $M/M/c$ queues the queue length can be found using the following \citep{ross2006introduction}:

\begin{equation} \label{eq:4.6}
    L_q = \frac{P_0 (\frac{\lambda}{\mu})^c \rho}{c! (1-\rho)^2}
\end{equation}

where $P_0$ in \eqref{eq:4.6} is the probability of zero travellers in the system, otherwise known as the Erlang C formula \citep{kleinrock1975queueing, ross2006introduction}:

\begin{equation} \label{eq:4.7}
    P_0 = \frac{1}{1 + (1-\rho) \frac{c!}{(c\rho)^c}\sum^{c-1}_{m=0} \frac{(c\rho)^m}{m!}}
\end{equation}

As observed in equation \eqref{eq:4.7}, any terms including the number of agents $c$, such as the factorial and power terms produce extremely large numbers as $c$ becomes large. As such, we define a transformation which assists the calculation of $P_0$ with large numbers. To do so, we set function $f(c,\rho)$ as follows:

\begin{equation} \label{eq:4.8}
    f(c,\rho) = \frac{c!}{(c\rho)^c}\sum^{c-1}_{m=0} \frac{(c\rho)^m}{m!}
\end{equation}

We note that equation \eqref{eq:4.8} contains terms of the cumulative distribution function (CDF) of a Poisson distribution. As such, we transform equation \eqref{eq:4.8} as follows:

\begin{equation} \label{eq:4.9}
    f(c,\rho) = \frac{c!}{(c\rho)^c} e^{c\rho} F_{Poisson}(c-1, c\rho)
\end{equation}

Where $F_{Poisson}(c-1, c\rho)$ in \eqref{eq:4.9} is the CDF value of a Poisson distribution with mean $c\rho$ and cumulative probability $P(X \leq c-1)$. Using exponents of natural logarithms and the gamma function\footnote{For a Gamma function $\Gamma(n)$ we have $\Gamma(n)=(n-1)!$.}, we have the following form for $f(c,\rho)$:

\begin{equation} \label{eq:4.10}
    f(c,\rho) = \exp\big[\ln(\Gamma(c+1)) - c \ln(c\rho) +c\rho\big] F_{Poisson}(c-1, c\rho)
\end{equation}

Consequently we use the function $f(c,\rho)$ in equation \eqref{eq:4.10} to transform function $P_0$ to a more convenient version for large values of $c$, as shown below:

\begin{equation} \label{eq:4.11}
    P_0 = \frac{1}{1+(1-\rho)f(c,\rho)}
\end{equation}

The average wait time in the queue $W_q$ models the wait time until assignment. As such, the wait time from assignment to pickup is embedded in the service time. Nonetheless, more vehicles in the fleet would produce less pickup wait times on average. That is, assuming vehicles and requests are uniformly distributed in space. Consequently, by embedding the pickup wait time in the service time, we cannot regard the service time as a constant (i.e. mean of an exponential distribution).

Instead, we choose to model pickup wait time by exploiting the spatial characteristics of the underlying network. By assuming all available vehicles are uniformly distributed in space, we divide the total network area $A$ by the number of idle vehicles $V$. As such, each vehicle has a coverage area of $\frac{A}{V}$. For a single idle vehicle with velocity $v$, we can identify the pickup time of an assigned request using $t_p =\frac{x}{v}$. Where $x$ is the distance from the location of the available vehicle to the pickup spot. 

Assuming the number of idle vehicles can vary, we can define $t_p$ using the average velocity in the network $\Bar{v}$ and a derived average pickup distance. If an incoming request can be located in any of the coverage areas $\frac{A}{V}$ in the network, then we need to define the average distance between a request location and any available vehicle in the fleet. To do so, we assume that each vehicle covers a square area\footnote{This error of this approximation reduces with increasing vehicle availability.}. Since requests and available vehicles are uniformly distributed in the network, we need to find the average distance between any two points in a square coverage area $\frac{A}{V}$. However, to account for non-uniform dispersion of vehicles across area $A$, we introduce factor $\psi \in (0, 1]$, such that the coverage area is transformed to $\frac{A}{\psi V}$.

Assuming the coverage areas to be squares, we consider the average distance between two uniformly distributed points in a square domain. \cite{burgstaller2009average} proved that the distance between any two such points in a square is approximately equal to $0.52\sqrt{a}$, where $a$ is the square's side length. By considering a factor $\phi$ analogous to taxicab geometry, we derive the average distance between a request and total available vehicles $V$ in a network with area $A$ is equal to $0.52 \phi \sqrt{\frac{A}{\psi V}}$. We identify the value of $\phi$ by considering the average ratio of a network path over the haversine distance of between two road nodes. Consequently, we define the average pickup wait time $t_p(V)$ for requests using the following equation:

\begin{equation} \label{eq:4.12}
    t_p(V) = \frac{0.52 \phi}{\Bar{v}} \sqrt{\frac{A}{\psi V}} \quad \forall V > 0
\end{equation}

The pickup wait time derived in equation \eqref{eq:4.12} is in line with estimates derived in the works of \cite{zha2018geometric} and \cite{korolko2018dynamic}. Identifying the variation of $V$ with respect to the fleet size $c$ is a non-trivial task due to a variation of $V$ with time after changes in parameters such as the rate of request arrivals or the average velocity during specific hours. To understand the nature of the fluctuation, we can consider how $V$ changes on each frame in time. By assuming a discrete-time step $t$ and a set of time steps $T$, the number of idle vehicles $V_t$ at time step $t$ is governed by the following equation:

\begin{equation} \label{eq:4.13}
    V_t = V_{t-1} + V_{t}^{I} - V_{t}^{O} \quad \forall t \in T
\end{equation}

Where $V_{t}^{I}$ and $V_{t}^{O}$ in equation \eqref{eq:4.13} represent the incoming idle vehicles and the newly outgoing/occupied vehicles respectively at time step $t$. We can regard the value of $V_{t}^{O}$ as the constant value of traveller arrivals per time step $t$ over the period $T$; however, the value of incoming idle vehicles depends, not only on the constant average trip time but also on the value of the average pickup wait time. Furthermore, $V_{t}^{I}$ at time step $t$ is decided by the average pickup wait time and by extension the number of idle vehicles of previous time steps $t'$. 

As an example, consider a per-minute time step $t$, where the average trip time is 10 minutes, and the average pickup wait time at $t_0$ is 2 minutes. With these values, all the outgoing vehicles at $t_0$ would be added to the incoming idle vehicles of time step $t_0 + 12$. If the pickup wait time at $t_0 + 1$ is reduced to 1 minute with the trip time remaining constant, then the outgoing vehicles $V_{t_0 + 1}^{O}$ would also be added to the incoming idle vehicles of $t_0 + 12$. Considering this complex behaviour, we could identify the value of minimum fleet size $c_0$, either by using non-linear integer programming or by assuming a continuous representation of idle vehicles $V$ and time $t$. In both scenarios, we would seek to identify the lowest value of $V_0$ (which we can set to $c$), for which equation \eqref{eq:4.13} is still feasible across all time steps $t$ with $V_{t}^{O}$ assumed to be constant.

As a consequence of our model, the minimum required fleet size $c_0$ is larger if we account for pickup wait time. Furthermore, since the form of $P_0$ depends on $\rho$, changes on the structure of utilization $\rho$ must be reflected on equations \eqref{eq:4.6} and \eqref{eq:4.7} as well. Nonetheless, the derivation of such formulas is non-trivial and beyond the scope of our ABM framework. However, will see in section \ref{sec: Discussion} that due to the size of fleets in practical implementations, the $M/M/c$ queue model converges to $M/M/\infty$, which implies no waiting in the queue (assignment wait).

Since the value of $c_0$ is governed by the value $V_0$, we expect that for $c_0$, $V(t)$ would initially drop to a minimum, and the reach a steady-state value as idle vehicles start returning from trips. Consequently, the value of idle vehicles $V$ in equation \eqref{eq:4.12} for pickup wait time $t_p(V)$ could be greater than $V_{t}^{O}$ at the critical fleet size case if the steady state is above the minimum. However, for $c < c_0$, we expect the value of $t_p(V)$ to be equal to the maximum, with $V$ equal to $V_{t}^{O}$. This would imply a possible discontinuity in the measurement of pickup wait times $t_p$ and utilization $\rho$ for $c < c_0$ and $c \geq c_0$.

By assuming the service rate introduced in equation \eqref{eq:4.1} takes into account the steady-state pickup wait time $\bar{t}_p$, we can express $\mu$ using the following equation:

\begin{equation} \label{eq:4.15}
    \mu = \frac{1}{\bar{t} + \bar{t}_p}
\end{equation}

As such, the fleet utilization rate $\rho$ in equation \eqref{eq:4.1} transforms to the following equation:

\begin{equation} \label{eq:4.16}
    \rho = \frac{\lambda}{c} (\bar{t} + \bar{t}_p)
\end{equation}

The discontinuity in pickup wait times $t_p$ at $c_0$, prohibits the identification of $c_0$ by setting the utilization $\rho$ in equation \eqref{eq:4.16} equal to 1. 

\section{Discussion} \label{sec: Discussion}
In section \ref{sec: Queue}, we modelled the function of a ride-sourcing fleet as a queue of multiple servers to identify the critical size of the fleet. However, by analysing the effects of the variation of pickup wait times with idle vehicles in time, we deduced that identifying a minimum value for fleet size is an increasingly complex problem. As such, we resort to the use of our ABM framework from section \ref{sec: Framework} to identify the minimum fleet size for ride-sourcing operations under specific parameters.

\subsection{Model Structure} \label{sec: ModelStructure}

We structure an ABM with the three basic types of agents highlighted in Table \ref{tab: agents}. For the traveller and vehicle agents, we chose the basic stage logic outlined in Table \ref{tab: States}, but we omit the Aborted state for the traveller agents. Instead, we assume that travellers never abort the service so that we adequately capture the relationships highlighted in Section \ref{sec: Queue}. We also define a single active state for the operator agent. We, therefore, use a simplified model where travellers submit requests for private trips at the operator at their request times and are assigned to their closest idle vehicle using a first-in-first-out (FIFO) assignment routine. Travellers exit the system once they are served by a vehicle. The operator also routes vehicles through the network using the $A^*$ algorithm.

To identify a value for minimum fleet size relevant to the models in Section \ref{sec: Queue}, we ignore complex behaviours such as mode choice, charging and parking station choice, pricing and fleet management strategies. The inclusion of a mode choice model would defeat the purpose of identifying unstable queues as travellers would always choose a different mode if they are subjected to long wait times. Electric charging behaviour is expected to slightly increase the minimum fleet size during a simulation hour due to more unavailable vehicles, but that is solely dependent on the choice of battery and the velocity fluctuation. Furthermore, we assume it would be reasonable to use a parking selection and fleet management model if the minimum required fleet size was known in advance.

To initiate and perform the simulation, we first instantiate non-agent (road-network) and agent objects in the simulator. We note that this process does not refer to the model/code building procedure described in Section \ref{sec: Build} but refers to the prerequisite step of initiating instances of all objects required for the simulation to run. During initialization, we create the road network $G$, the set of agent types $A$ and the set of simulation time steps $T$. We note that the set of travellers $R$ in $A$ is an empty set at the end of the initialization, as travellers are generated during the simulation time. 

In each time step $t$ of the simulation, initially, we add any new traveller requests. Then, the simulator iterates through all the agent sets to progress the state logic of each agent. The FIFO assignment is performed as part of the logic step of the operator, whereas the $A^*$ routing algorithm is performed by any vehicle which enters a travelling state such as Travelling to origin or Travelling to destination (Table \ref{tab: States}). Furthermore, agent KPIs and properties are updated at each time step of the simulation. To achieve an acceptable precision in the representation of our agent-based model, we set the time step size $t$ to be one second.

\subsection{Model Instances} \label{sec: 6.2}
\begin{figure}[h]
\centering
\includegraphics[width=0.5\textwidth]{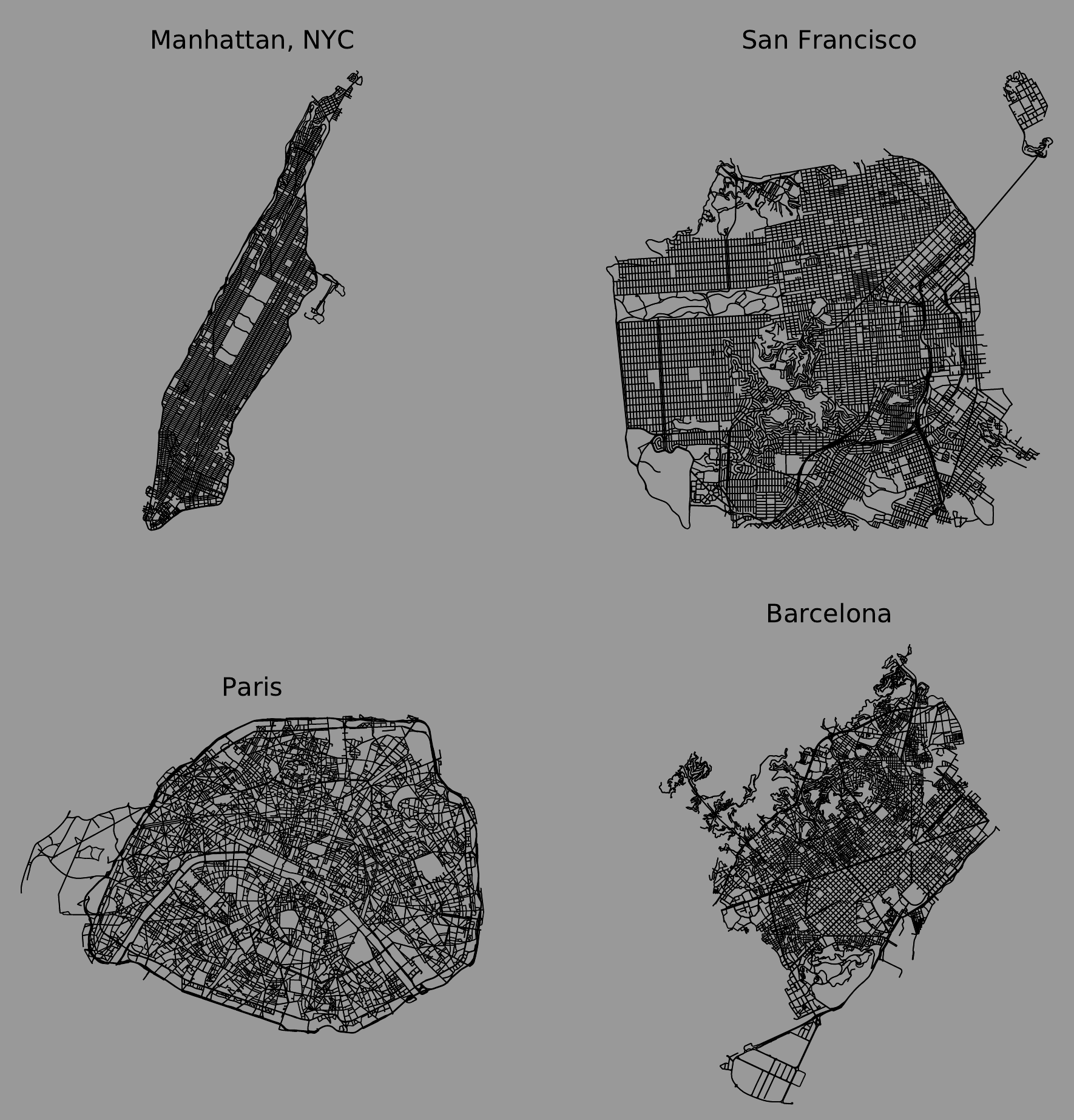}
\caption{Road network geometries for urban areas used in the analysis.}
\label{fig:6.1}
\end{figure}

To test our methodology, we selected four urban areas to create case study networks; namely the island of Manhattan in New York City, and the areas within the city boundaries of San Francisco, Paris and Barcelona. The underlying road networks and link travel times were obtained using the OSMnx library \citep{BOEING2017126}. We omitted endogenous congestion in our model by assuming a small proportion of traffic accounts to ride-sourcing vehicles. However, we applied a 20\% reduction to the free-flow speeds in residential and motorway link segments, and 40\% elsewhere, to account for exogenous congestion. Figure \ref{fig:6.1} shows the extends and the geometries of the road-networks used for the analysis.

Using the dataset provided by \cite{TLC2019} and the \cite{SanFranciscoCountyTransportationAuthority2020}, we created typical demand profiles for morning peak hours in Manhattan and San Francisco, respectively. While the trip dataset for Manhattan already included trip times, the San Francisco dataset only included inbound and outbound zonal trip counts. As such, to generate a synthetic zone Origin-Destination (OD) matrix for San Francisco, we used the Iterative Proportional Fitting (IPF) algorithm \citep{10.2307/2983403}. 

Using the zone polygons provided by \cite{SanFranciscoCountyTransportationAuthority2020}, the San Francisco road-network and the resulting OD matrix, we sampled random OD nodes within the zone polygons for each OD zone pair. In the absence of ride-sourcing trip datasets for the cities of Paris and Barcelona, we randomly sampled origin and destination node pairs for 10000 and 7000 trips per hour respectively. To achieve a realistic dispersion of the trips in time for each city other than Manhattan\footnote{Trip times for Manhattan are available in the dataset \citep{TLC2019}}, we sampled inter-arrival times using an exponential distribution for three hours. Table \ref{tab: instances} outlines the properties of model instances used to test the simulation.

\begin{table}[]
\centering
\caption{Descriptions of urban area instances used in the analysis.}
\label{tab: instances}
\begin{tabular}{@{}lccccc@{}}
\toprule
\begin{tabular}[c]{@{}c@{}}Instance\\ {}\end{tabular} &
  \begin{tabular}[c]{@{}c@{}}$A$\\ {[}$km^2${]}\end{tabular} &
  \begin{tabular}[c]{@{}c@{}}$\phi$ \\ {}\end{tabular} &
  \begin{tabular}[c]{@{}c@{}}$\Bar{v}$\\ {[}$km/h${]}\end{tabular} &
  \begin{tabular}[c]{@{}c@{}}$\lambda$\\ {[}/h{]}\end{tabular} &
  \begin{tabular}[c]{@{}c@{}}$\Bar{t}$\\ {[}h{]}\end{tabular} \\ \midrule
Manhattan     & 59.1  & 1.36 & 24.5 & 11607 & 0.11 \\
San Francisco & 121.4 & 1.30 & 25.4 & 9454  & 0.20 \\
Paris         & 105.4 & 1.31 & 23.2 & 10000 & 0.27 \\
Barcelona     & 101.9 & 1.41 & 24.5 & 7000  & 0.28 \\ \bottomrule
\end{tabular}
\end{table}

\subsection{Analysis} \label{sec: 6.3}
To identify the critical fleet size for each instance, we performed multiple simulations of the ride-sourcing service, varying the fleet size in each simulation run. In doing so, we recorded the length of the FIFO list on each time step, which is interpreted as the queue length introduced in Section \ref{sec: Queue}. We regard any instances with fleet sizes that produce departures from a constant queue length as unstable and below the critical fleet size.

\begin{figure}[h]
\centering
\includegraphics[width=0.7\textwidth]{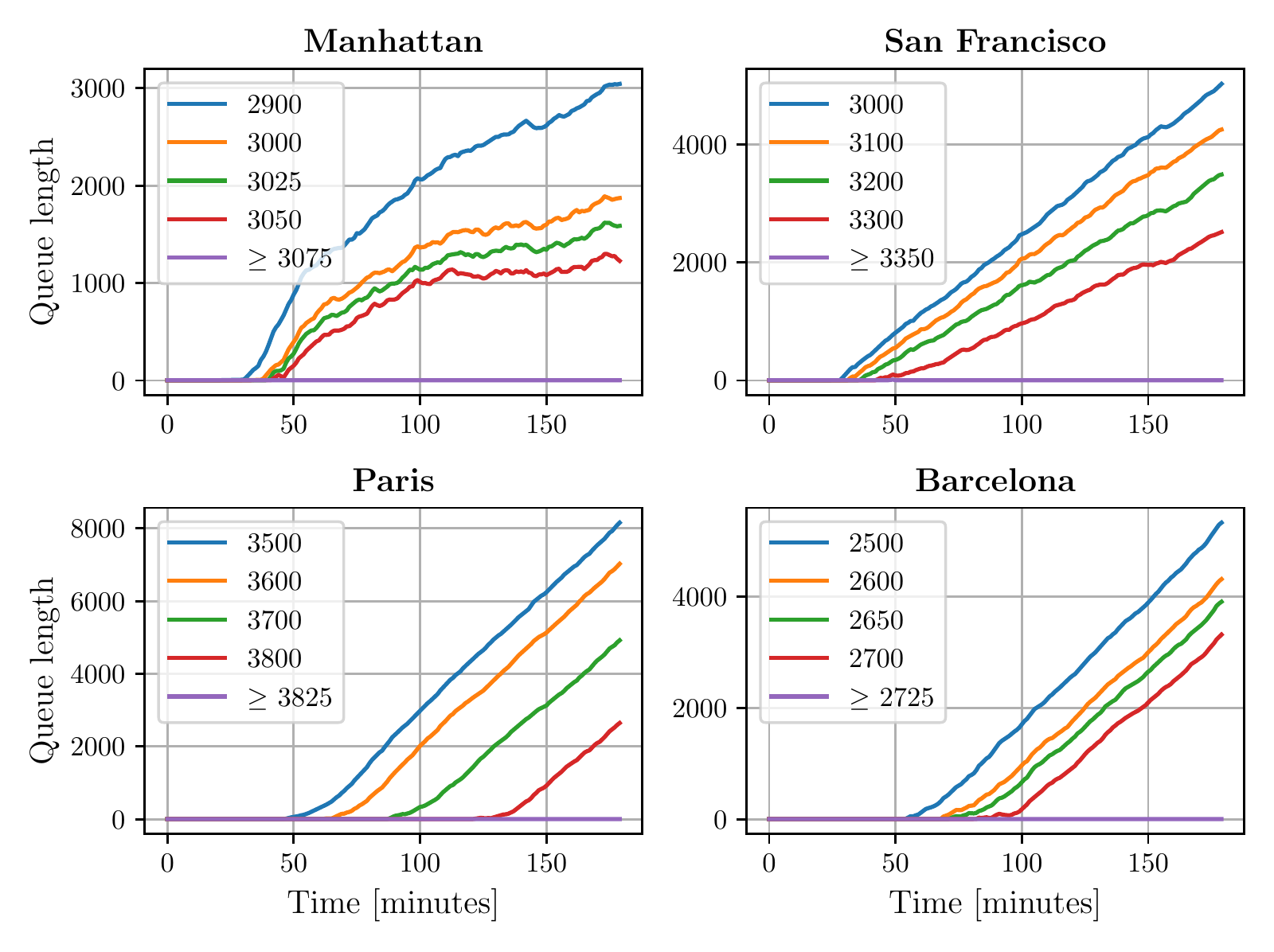}
\caption{Queue lengths for each instance over the simulation period.}
\label{fig:6.2}
\end{figure}

Observing figure \ref{fig:6.2} for each modelled city and instance, we deduce as expected, that if the fleet size is greater than the critical fleet size, the average queue length has the same constant value for all fleet sizes. This constant value is equal to the number of travel requests per second, which implies that there is no wait in the queue and travellers are assigned to vehicles immediately. 

The zero queue wait for stable queues is a direct implication of the volume of requests per hour $\lambda$. Revisiting equation \eqref{eq:4.13}, we see that even for $V_{t-1} = 0$, due to constant non-zero traveller arrivals per time step $t$, the value of $V_{t}^I$ must be at least equal to the traveller arrivals per second, for stability. The required abundance of idle vehicles is also expected when using equations \eqref{eq:4.1}-\eqref{eq:4.11} by assuming a converged low-value pickup the pickup-wait time. This directly implies that our $M/M/c$ queue model converges to an $M/M/\infty$ model for $\rho \geq 1$. 

\begin{figure}[h]
\centering
\includegraphics[width=0.7\textwidth]{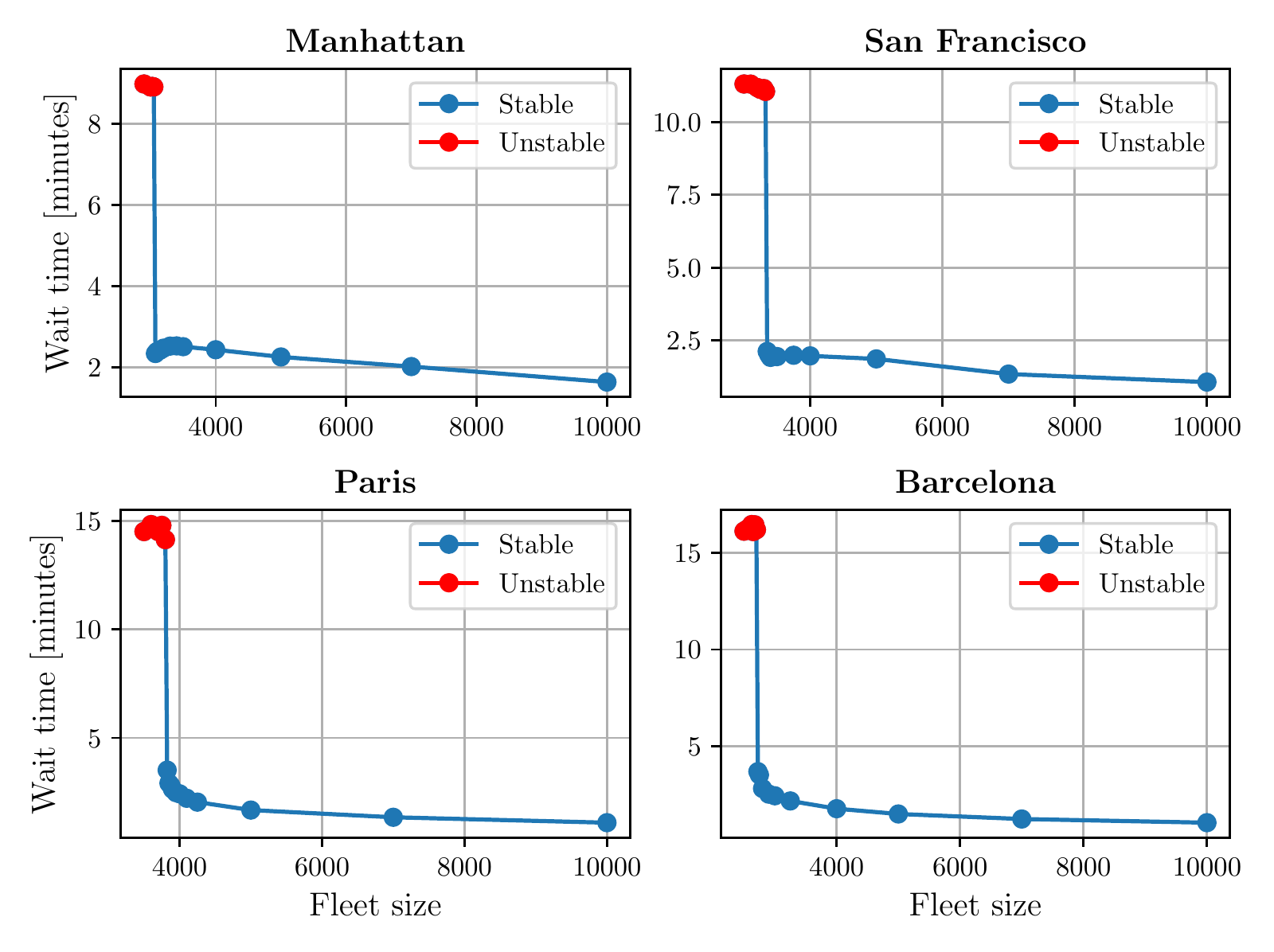}
\caption{Pickup wait time variation for each instance.}
\label{fig:6.3}
\end{figure}

To further scrutinize the shift from unstable to stable queues, we also recorded the pickup wait times of each served traveller in the simulations as shown in \ref{fig:6.3}. To achieve steady-state pickup wait times in all cases, we only record pickup wait times in the hours after the queue becomes unstable (see Figure \ref{fig:6.2}). In Table \ref{tab: max pickup} we provide estimates of maximum pickup wait times using Equation \eqref{eq:4.12} and the values from Table \ref{tab: instances} for $\psi=1$ (homogeneous dispersion). We estimate the number of idle vehicles $V$ in each case, to the value of incoming requests per second, transformed from $\lambda$ in Table \ref{tab: instances}. 

In that respect, we observe in Table \ref{tab: max pickup}, that the maximum simulation pickup wait times for unstable queues are not always closely approximated by our pickup wait time model in equation \eqref{eq:4.12}. Inaccuracies in the calculation might be down to origin locations dispersion (for Manhattan and San Francisco) and shape irregularities (network areas are not square). Nonetheless, we deduce that with appropriate calibration of parameters to minimize deviation from observed behaviour, the model can serve as a reasonable approximation in aggregate optimization models.  

\begin{table}[h]
\centering
\caption{Maximum pickup wait time values using approximation and simulation.}
\label{tab: max pickup}
\begin{tabular}{@{}lcccc@{}}
\toprule
Instance &
  $V$ &
  \begin{tabular}[c]{@{}c@{}}$t_p$ (Eq\eqref{eq:4.12})\\ {[}min{]}\end{tabular} &
  \begin{tabular}[c]{@{}c@{}}$t_p$ (Simulation)\\ {[}min{]}\end{tabular} &
  \begin{tabular}[c]{@{}c@{}}Absolute Error\\ {[}\%{]}\end{tabular} \\ \midrule
Manhattan     & 3.28 & 7.35  & 8.43  & 12.8 \\
San Francisco & 2.63 & 10.85 & 10.71 & 1.3  \\
Paris         & 2.78 & 10.84 & 14.15 & 23.4 \\
Barcelona     & 1.94 & 13.01 & 15.75 & 14.4 \\ \bottomrule
\end{tabular}
\end{table}

\begin{figure}[H]
\centering
\includegraphics[width=0.7\textwidth]{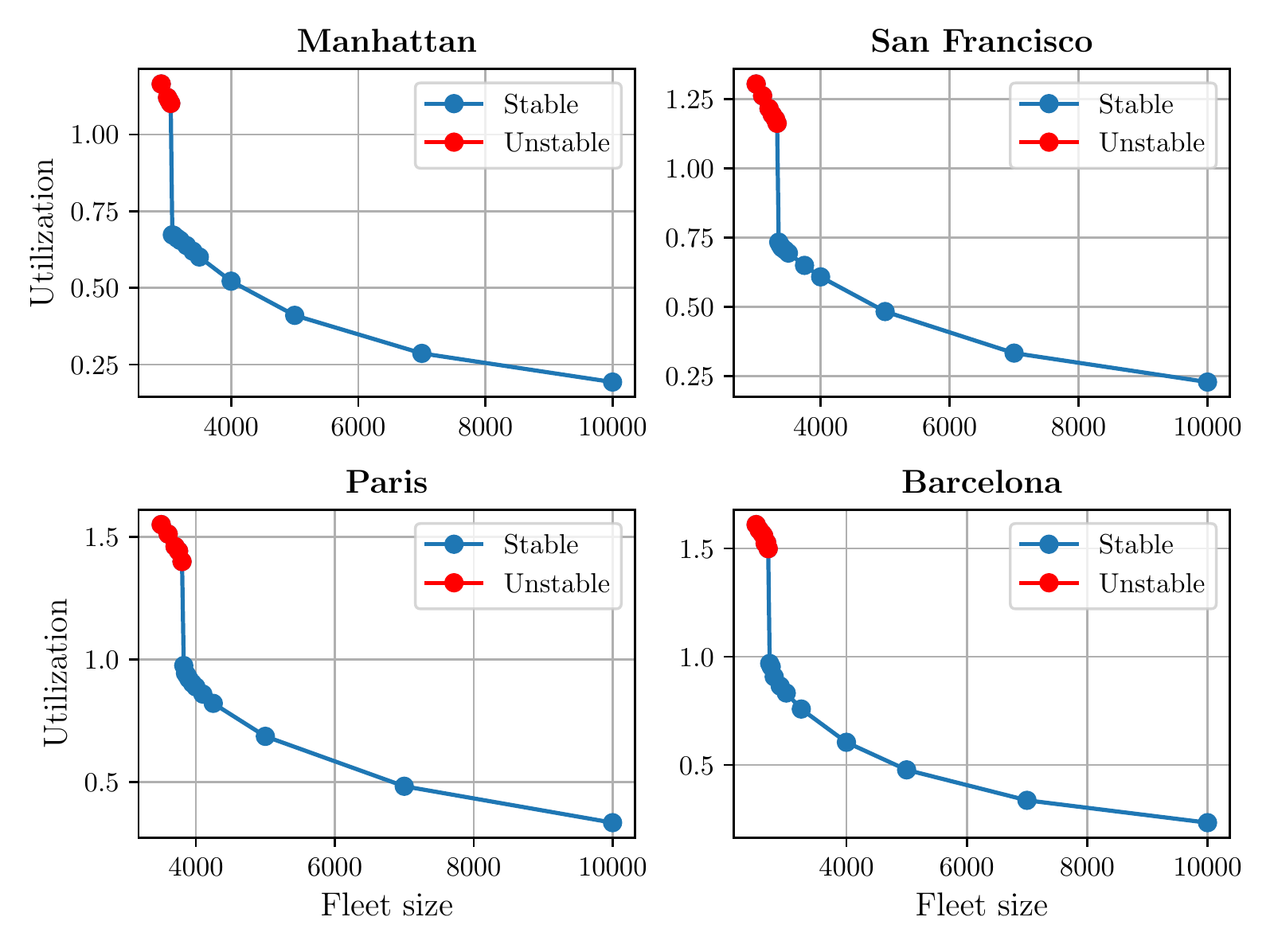}
\caption{Utilization variation for each instance.}
\label{fig:6.4}
\end{figure}

Using the average trip times and pickup wait times of each instance, we were also able to calculate the fleet utilization as shown in equation \eqref{eq:4.16}. Similarly to the wait times, in figure \ref{fig:6.4}, in all models, we identify a jump in values from unstable to stable queues. This discontinuity in pickup wait times and fleet utilization is expected and attributed to the fluctuation of idle vehicles in the simulation, as explained in Section \ref{sec: Queue}.

\begin{figure}[H]
\centering
\includegraphics[width=0.63\textwidth]{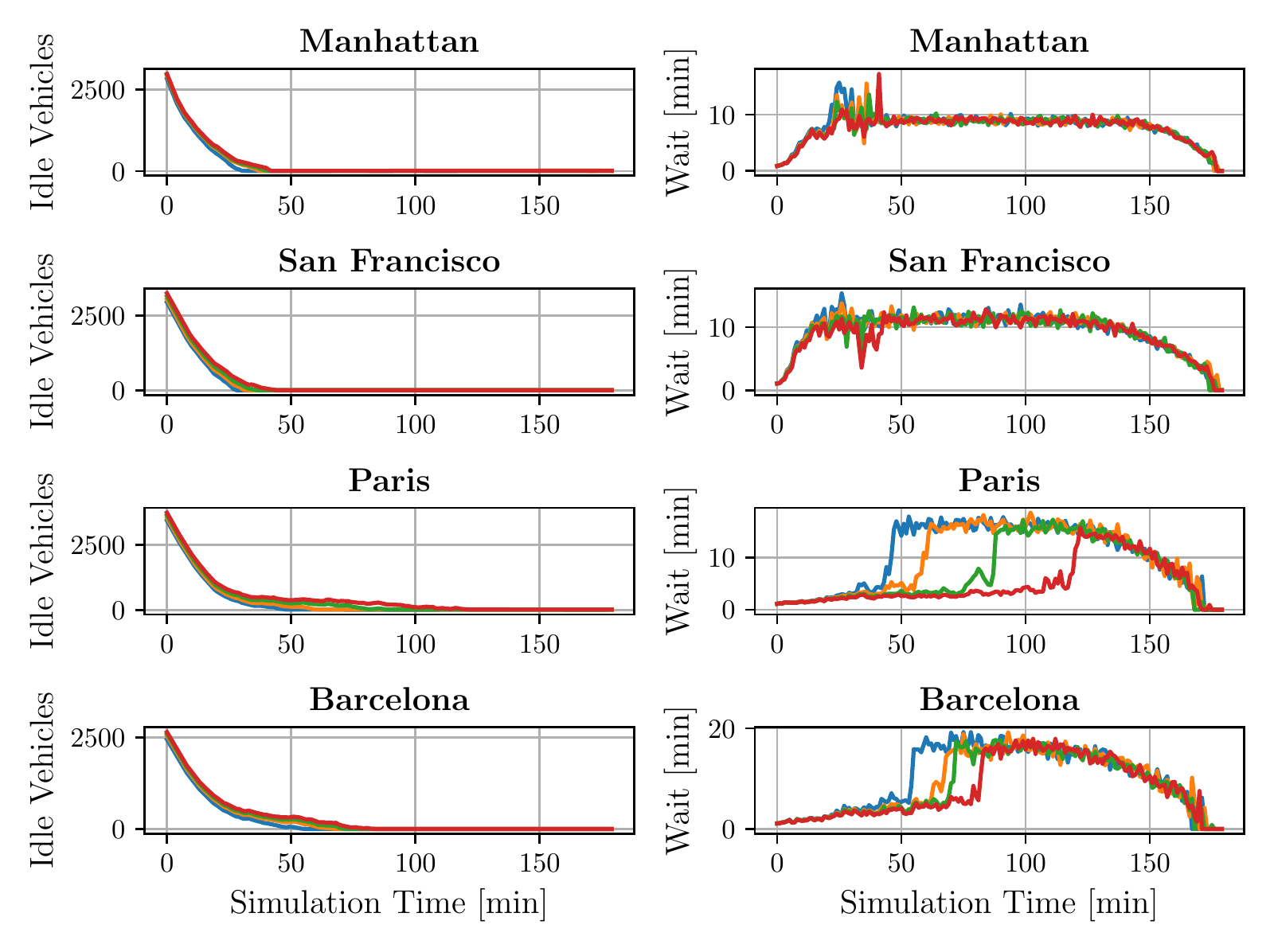}
\caption{Variation of idle vehicle counts and pickup wait time over unstable queue settings.}
\label{fig:6.5}
\end{figure}

\begin{figure}[H]
\centering
\includegraphics[width=0.63\textwidth]{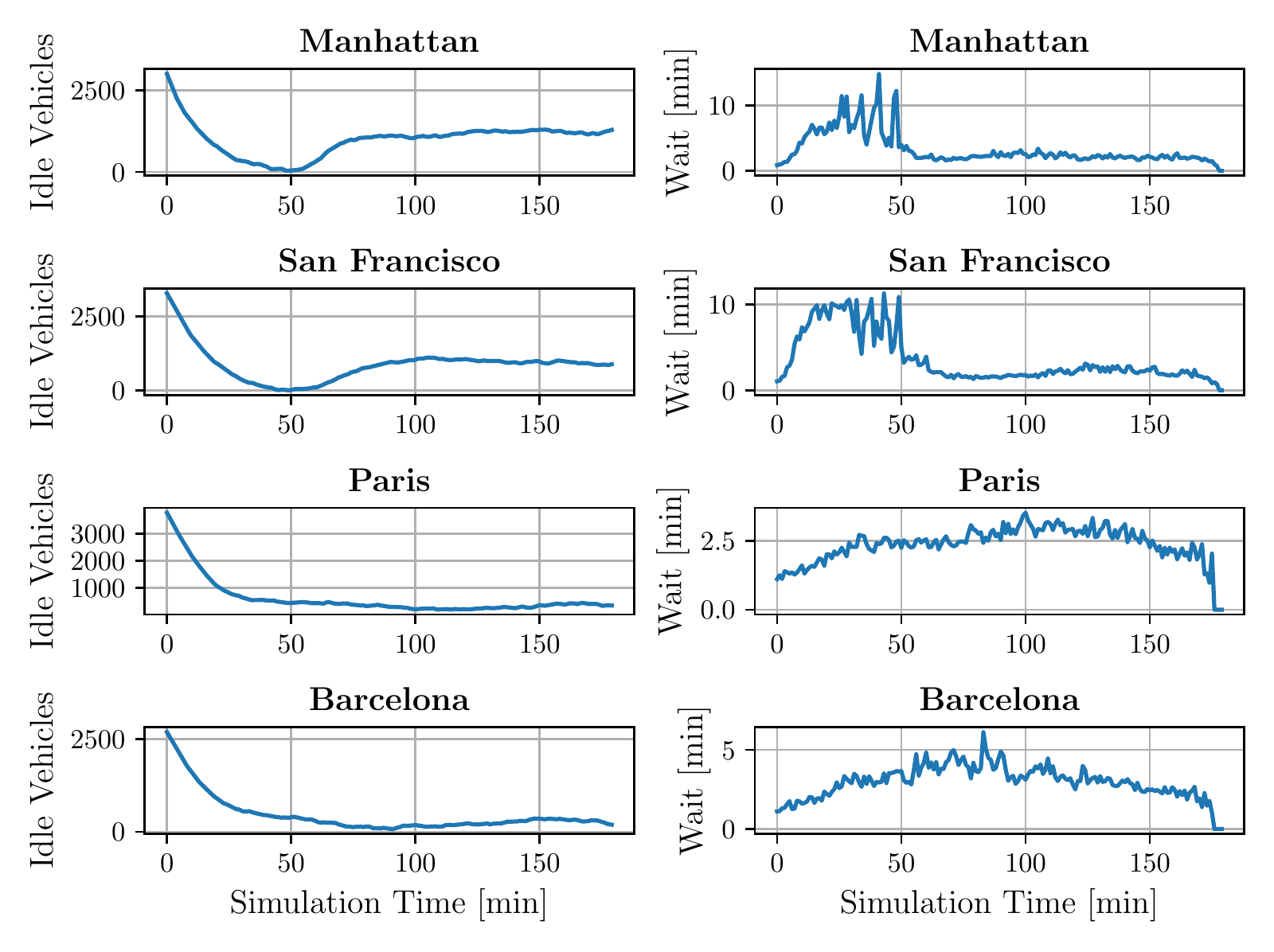}
\caption{Variation of idle vehicle counts and pickup wait time over a stable queue setting.}
\label{fig:6.6}
\end{figure}

In figures \ref{fig:6.5} and \ref{fig:6.6}, we outline the fluctuation of idle vehicles and pickup wait times experienced by travellers for cases of unstable and a case of a stable queue respectively for all cities. We see that in the case of unstable queues (figure \ref{fig:6.5}), the value of idle vehicles decreases at an exponential rate and never recovers after reaching the minimum. When observing figure \ref{fig:6.6}, for the case of the stable queue, as expected, after an initial dip, the idle vehicles stabilize to a constant rate. The opposite trend is identified for pickup wait time, with the wait time reaching a maximum as the idle vehicles decrease and thereby also reaching a steady state in line with a constant flow of idle vehicles. The above observations are more prominent the instances of Manhattan and San Francisco.

\subsection{Model Validity} \label{sec: 6.4}
Through sections \ref{sec: ModelStructure} to \ref{sec: 6.3}, we structured a bespoke ABM using the framework proposed in section \ref{sec: Framework} to identify minimum AV TNC fleet sizes for various networks. In this section we aim to scrutinize the validity of our model and discuss how our ABM framework could have been used to solve different problems. Our model, although simplistic, offers a swift introduction on ABM structuring and a solid starting implementation for investigating AV TNC system dynamics.

In section \ref{sec: ModelStructure}, we assumed no traveller acumen or competition in the market. As a consequence, travellers never abort the service and the requests pile up in the queue if the fleet size is unsustainable. The objective of this simplification is to acquire an upper bound on the minimum fleet size in each network implementation. Additionally, we did not account for endogenous congestion in the system, as we considered fleet sizes which are only a negligible fraction of the background traffic in the network.

In realistic scenarios, intelligent travellers with wait times beyond the ones they deem acceptable, would abort the service and resort to other solutions, thereby not contributing to long unsustainable queues. Furthermore, an AV TNC would increase ride prices during peak travel via the use of dynamic pricing strategies, to sway excess demand to other modes and ensure a stable service. This system response would also imply complex and intelligent traveller and operator agents in the ABM.

Nonetheless, even in the absence of such complicated ABM scenarios discussed above, the minimum sustainable fleet size results found in the previous section do serve as upper bounds for the service. Furthermore, the model used serves as an extensible basis for more complicated problems, such as optimal pricing strategies, assignment operations and fleet management decision making (i.e. electric charging, parking, maintenance, idle redistribution).

\begin{figure}[t]
\centering
\includegraphics[width=1\textwidth]{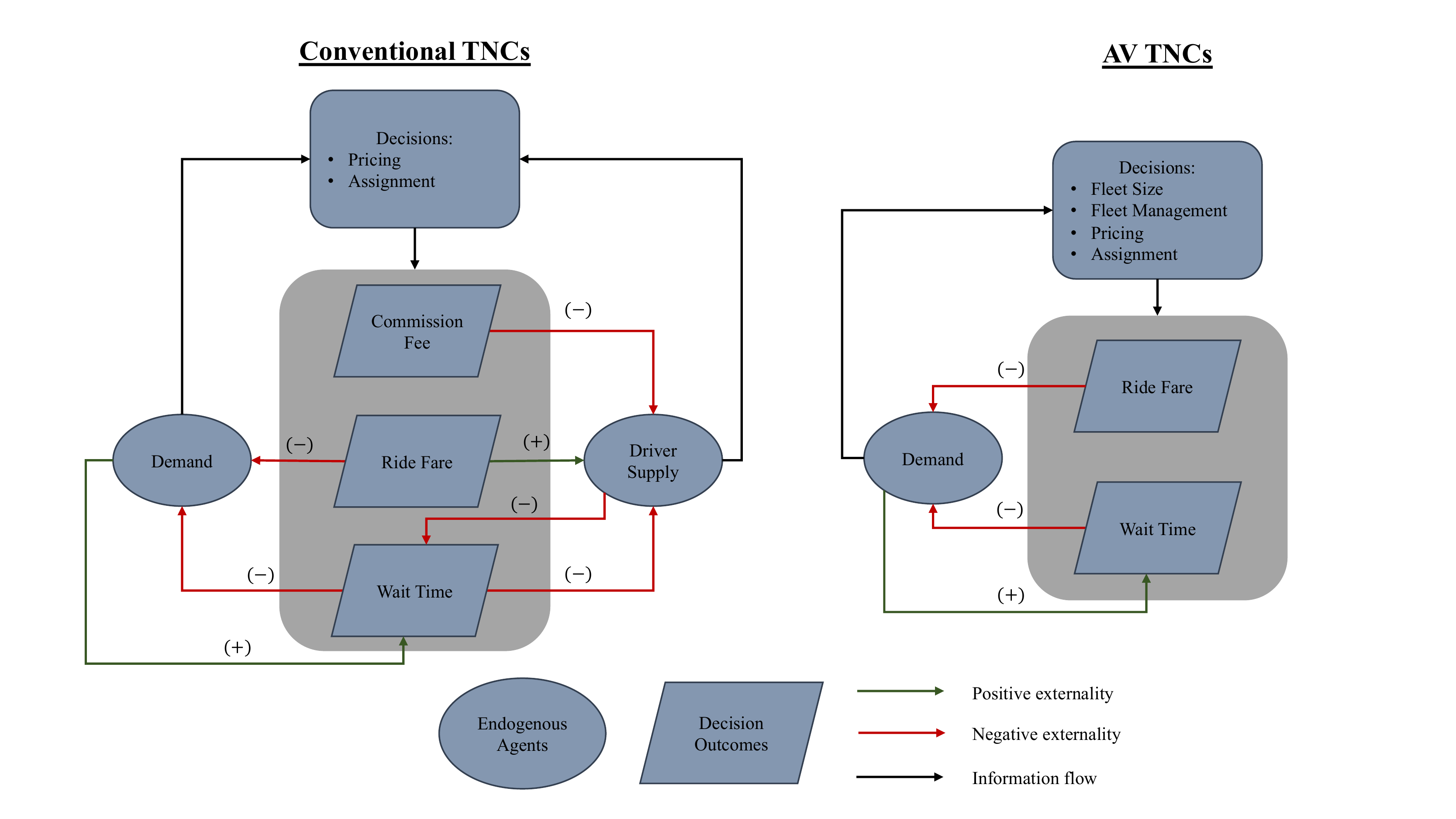}
\caption{Conceptualization of relationships in conventional and AV TNC operations identified across the literature.}
\label{fig:6.7}
\end{figure}

To comprehend the required extensible features for the basic ABM structure to be used in more complicated problems, we first identify the relationships of the different operations in the AV TNC system, and what the differences are in the switch from conventional ride-sourcing platforms to autonomous services. Using the comprehensive literature reviews on ride-sourcing platforms and SAV (similar to AV TNCs) operations by \cite{Wang2019} and \cite{narayanan2020shared} respectively, we conceptualize the relationships in both conventional and AV ride-sourcing using the diagrams in figure \ref{fig:6.7}.

In conventional TNC systems, both demand for rides and supply of drivers is endogenous, with travellers and drivers making online decisions depending on wait time and pricing specifics. The ride-sourcing platform requires a good understanding of both sides, so as to set system parameters which will lead to traveller/driver behavior that will maximize revenue or profits. In the case of AVs, TNCs have complete control of the supply and need to choose their strategies so as to efficiently attract and serve demand.

As a consequence of the above description, the most important extension in our ABM model to tackle more complicated problems, is intelligent traveller behaviour. We regard this extension as essential, especially in testing assignment and pricing strategies which are heavily reliant on traveller choice. Furthermore, realistic implementations of traveller behavior would require choice models calibrated using TNC ride data or stated preference surveys. Additional extensions could be endogenous congestion and the use of a car following model in vehicle movement. Nonetheless, these extensions only become relevant when the fleet size is comparable to the background traffic of a network, so as to have a sizeable impact on the link velocities. 

\section{Conclusion} \label{sec: Conclusion}

In this paper, we conveyed the use of agent-based models for simulating ride-sourcing operations by researchers, operators and policymakers alike. In reviewing the relevant literature, we outlined the need for a prescribed approach in designing such agent-based models. We then proposed a framework for ride-sourcing simulations founded from the fundamentals of agent-based modelling and suggested a modular approach for creating bespoke simulations using our framework with the aid of object-oriented programming. To showcase the necessity of our agent-based modelling framework and outline the complexity of ride-sourcing systems, we considered the validation of an aggregated queuing theoretical model of a ride-sourcing platform, to tackle the problem of minimum fleet size.

To test the validity of our queuing theoretical model and highlight the limits of aggregated representations of the system across a variety of road-network structures, we selected four case study applications; the urban areas of Manhattan, San Francisco, Paris and Barcelona. The results imply $M/M/\infty$ queues govern the operation of the ride-sourcing service across the entire urban network for the selected demand volumes. We also justify that the inclusion of pickup waiting time in modelling ride-sourcing operations via queuing theory is essential in the identification of the minimum fleet size. Finally, we conclude that the use of agent-based modelling is required to fully capture the complex dynamics which govern the critical fleet sizes in ride-sourcing systems.

Our suggestions for future research and continuity of our work span across multiple levels; currently, the two most common assumptions in modelling include no endogenous implication on network velocities by ride-sourcing fleets due to small sizes compared to background flows, or assuming the entire traffic is composed of discrete vehicles as agents in the model. We, therefore, believe that the development of a unified framework for discrete and continuous network flows would improve the accuracy of any ride-sourcing operation simulator. Furthermore, our model is limited by not using sophisticated assignment and idle vehicle redistribution methods. Such methods could have a sizeable impact in reducing vehicle mileage, and traveller pickup wait times, thereby future studies could investigate their effects on the minimum fleet size. Finally, our omission of traveller acumen to facilitate the proof of concept could be relaxed, also with more complicated market structures, to investigate how intelligent traveller behaviour can induce emergent effects in the modal split of a city.

\bibliography{References}

\begin{thebibliography}{62}
\expandafter\ifx\csname natexlab\endcsname\relax\def\natexlab#1{#1}\fi
\providecommand{\url}[1]{\texttt{#1}}
\providecommand{\href}[2]{#2}
\providecommand{\path}[1]{#1}
\providecommand{\DOIprefix}{doi:}
\providecommand{\ArXivprefix}{arXiv:}
\providecommand{\URLprefix}{URL: }
\providecommand{\Pubmedprefix}{pmid:}
\providecommand{\doi}[1]{\href{http://dx.doi.org/#1}{\path{#1}}}
\providecommand{\Pubmed}[1]{\href{pmid:#1}{\path{#1}}}
\providecommand{\bibinfo}[2]{#2}
\ifx\xfnm\relax \def\xfnm[#1]{\unskip,\space#1}\fi
%Type = Article
\bibitem[{Adam(2020)}]{adam2020special}
\bibinfo{author}{Adam, D.}, \bibinfo{year}{2020}.
\newblock \bibinfo{title}{Special report: The simulations driving the world's
  response to covid-19.}
\newblock \bibinfo{journal}{Nature} .
%Type = Inproceedings
\bibitem[{Alonso-Mora et~al.(2017)Alonso-Mora, Wallar and
  Rus}]{alonso2017predictive}
\bibinfo{author}{Alonso-Mora, J.}, \bibinfo{author}{Wallar, A.},
  \bibinfo{author}{Rus, D.}, \bibinfo{year}{2017}.
\newblock \bibinfo{title}{Predictive routing for autonomous mobility-on-demand
  systems with ride-sharing}, in: \bibinfo{booktitle}{2017 IEEE/RSJ
  International Conference on Intelligent Robots and Systems (IROS)},
  \bibinfo{organization}{IEEE}. pp. \bibinfo{pages}{3583--3590}.
%Type = Article
\bibitem[{Auld et~al.(2016)Auld, Hope, Ley, Sokolov, Xu and Zhang}]{Auld2016}
\bibinfo{author}{Auld, J.}, \bibinfo{author}{Hope, M.}, \bibinfo{author}{Ley,
  H.}, \bibinfo{author}{Sokolov, V.}, \bibinfo{author}{Xu, B.},
  \bibinfo{author}{Zhang, K.}, \bibinfo{year}{2016}.
\newblock \bibinfo{title}{{POLARIS : Agent-based modeling framework development
  and implementation for integrated travel demand and network and operations
  simulations}}.
\newblock \bibinfo{journal}{Transportation Research Part C: Emerging
  Technologies} \bibinfo{volume}{64}, \bibinfo{pages}{101--116}.
\newblock \DOIprefix\doi{10.1016/j.trc.2015.07.017}.
%Type = Article
\bibitem[{Balmer et~al.(2006)Balmer, Axhausen and Nagel}]{balmer2006agent}
\bibinfo{author}{Balmer, M.}, \bibinfo{author}{Axhausen, K.W.},
  \bibinfo{author}{Nagel, K.}, \bibinfo{year}{2006}.
\newblock \bibinfo{title}{Agent-based demand-modeling framework for large-scale
  microsimulations}.
\newblock \bibinfo{journal}{Transportation Research Record}
  \bibinfo{volume}{1985}, \bibinfo{pages}{125--134}.
%Type = Article
\bibitem[{Baptista et~al.(2016)Baptista, Farmer, Hinterschweiger, Low, Tang and
  Uluc}]{baptista2016macroprudential}
\bibinfo{author}{Baptista, R.}, \bibinfo{author}{Farmer, J.D.},
  \bibinfo{author}{Hinterschweiger, M.}, \bibinfo{author}{Low, K.},
  \bibinfo{author}{Tang, D.}, \bibinfo{author}{Uluc, A.}, \bibinfo{year}{2016}.
\newblock \bibinfo{title}{Macroprudential policy in an agent-based model of the
  uk housing market} .
%Type = Book
\bibitem[{Berlekamp et~al.(2018)Berlekamp, Conway and
  Guy}]{berlekamp2018winning}
\bibinfo{author}{Berlekamp, E.R.}, \bibinfo{author}{Conway, J.H.},
  \bibinfo{author}{Guy, R.K.}, \bibinfo{year}{2018}.
\newblock \bibinfo{title}{Winning ways for your mathematical plays}.
  volume~\bibinfo{volume}{1}.
\newblock \bibinfo{publisher}{CRC Press}.
%Type = Article
\bibitem[{Boeing(2017)}]{BOEING2017126}
\bibinfo{author}{Boeing, G.}, \bibinfo{year}{2017}.
\newblock \bibinfo{title}{Osmnx: New methods for acquiring, constructing,
  analyzing, and visualizing complex street networks}.
\newblock \bibinfo{journal}{Computers, Environment and Urban Systems}
  \bibinfo{volume}{65}, \bibinfo{pages}{126 -- 139}.
\newblock \URLprefix
  \url{http://www.sciencedirect.com/science/article/pii/S0198971516303970},
  \DOIprefix\doi{https://doi.org/10.1016/j.compenvurbsys.2017.05.004}.
%Type = Article
\bibitem[{Boesch et~al.(2016)Boesch, Ciari and Axhausen}]{boesch2016autonomous}
\bibinfo{author}{Boesch, P.M.}, \bibinfo{author}{Ciari, F.},
  \bibinfo{author}{Axhausen, K.W.}, \bibinfo{year}{2016}.
\newblock \bibinfo{title}{Autonomous vehicle fleet sizes required to serve
  different levels of demand}.
\newblock \bibinfo{journal}{Transportation Research Record}
  \bibinfo{volume}{2542}, \bibinfo{pages}{111--119}.
%Type = Article
\bibitem[{Bonabeau(2002)}]{bonabeau2002agent}
\bibinfo{author}{Bonabeau, E.}, \bibinfo{year}{2002}.
\newblock \bibinfo{title}{Agent-based modeling: Methods and techniques for
  simulating human systems}.
\newblock \bibinfo{journal}{Proceedings of the national academy of sciences}
  \bibinfo{volume}{99}, \bibinfo{pages}{7280--7287}.
%Type = Article
\bibitem[{Braun-Munzinger et~al.(2016)Braun-Munzinger, Liu and
  Turrell}]{braun2016agent}
\bibinfo{author}{Braun-Munzinger, K.}, \bibinfo{author}{Liu, Z.},
  \bibinfo{author}{Turrell, A.}, \bibinfo{year}{2016}.
\newblock \bibinfo{title}{An agent-based model of dynamics in corporate bond
  trading} .
%Type = Article
\bibitem[{Braverman et~al.(2019)Braverman, Dai, Liu and
  Ying}]{braverman2019empty}
\bibinfo{author}{Braverman, A.}, \bibinfo{author}{Dai, J.G.},
  \bibinfo{author}{Liu, X.}, \bibinfo{author}{Ying, L.}, \bibinfo{year}{2019}.
\newblock \bibinfo{title}{Empty-car routing in ridesharing systems}.
\newblock \bibinfo{journal}{Operations Research} \bibinfo{volume}{67},
  \bibinfo{pages}{1437--1452}.
%Type = Article
\bibitem[{Burgess et~al.(2013)Burgess, Fernandez-Corugedo, Groth, Harrison,
  Monti, Theodoridis, Waldron et~al.}]{burgess2013bank}
\bibinfo{author}{Burgess, S.}, \bibinfo{author}{Fernandez-Corugedo, E.},
  \bibinfo{author}{Groth, C.}, \bibinfo{author}{Harrison, R.},
  \bibinfo{author}{Monti, F.}, \bibinfo{author}{Theodoridis, K.},
  \bibinfo{author}{Waldron, M.}, et~al., \bibinfo{year}{2013}.
\newblock \bibinfo{title}{The bank of england’s forecasting platform:
  Compass, maps, ease and the suite of models}.
\newblock \bibinfo{journal}{documento de trabajo} .
%Type = Article
\bibitem[{Burgstaller and Pillichshammer(2009)}]{burgstaller2009average}
\bibinfo{author}{Burgstaller, B.}, \bibinfo{author}{Pillichshammer, F.},
  \bibinfo{year}{2009}.
\newblock \bibinfo{title}{The average distance between two points}.
\newblock \bibinfo{journal}{Bulletin of the Australian Mathematical Society}
  \bibinfo{volume}{80}, \bibinfo{pages}{353--359}.
%Type = Article
\bibitem[{Chang et~al.(2020)Chang, Harding, Zachreson, Cliff and
  Prokopenko}]{chang2020modelling}
\bibinfo{author}{Chang, S.L.}, \bibinfo{author}{Harding, N.},
  \bibinfo{author}{Zachreson, C.}, \bibinfo{author}{Cliff, O.M.},
  \bibinfo{author}{Prokopenko, M.}, \bibinfo{year}{2020}.
\newblock \bibinfo{title}{Modelling transmission and control of the covid-19
  pandemic in australia}.
\newblock \bibinfo{journal}{arXiv preprint arXiv:2003.10218} .
%Type = Article
\bibitem[{Chen and Kockelman(2016)}]{chen2016management}
\bibinfo{author}{Chen, T.D.}, \bibinfo{author}{Kockelman, K.M.},
  \bibinfo{year}{2016}.
\newblock \bibinfo{title}{Management of a shared autonomous electric vehicle
  fleet: Implications of pricing schemes}.
\newblock \bibinfo{journal}{Transportation Research Record}
  \bibinfo{volume}{2572}, \bibinfo{pages}{37--46}.
%Type = Article
\bibitem[{Chen et~al.(2016)Chen, Kockelman and Hanna}]{chen2016operations}
\bibinfo{author}{Chen, T.D.}, \bibinfo{author}{Kockelman, K.M.},
  \bibinfo{author}{Hanna, J.P.}, \bibinfo{year}{2016}.
\newblock \bibinfo{title}{Operations of a shared, autonomous, electric vehicle
  fleet: Implications of vehicle \& charging infrastructure decisions}.
\newblock \bibinfo{journal}{Transportation Research Part A: Policy and
  Practice} \bibinfo{volume}{94}, \bibinfo{pages}{243--254}.
%Type = Article
\bibitem[{Cleave et~al.(1995)Cleave, Brown and Payne}]{10.2307/2983403}
\bibinfo{author}{Cleave, N.}, \bibinfo{author}{Brown, P.J.},
  \bibinfo{author}{Payne, C.D.}, \bibinfo{year}{1995}.
\newblock \bibinfo{title}{Evaluation of methods for ecological inference}.
\newblock \bibinfo{journal}{Journal of the Royal Statistical Society. Series A
  (Statistics in Society)} \bibinfo{volume}{158}, \bibinfo{pages}{55--72}.
\newblock \URLprefix \url{http://www.jstor.org/stable/2983403}.
%Type = Incollection
\bibitem[{Crooks and Heppenstall(2012)}]{crooks2012introduction}
\bibinfo{author}{Crooks, A.T.}, \bibinfo{author}{Heppenstall, A.J.},
  \bibinfo{year}{2012}.
\newblock \bibinfo{title}{Introduction to agent-based modelling}, in:
  \bibinfo{booktitle}{Agent-based models of geographical systems}.
  \bibinfo{publisher}{Springer}, pp. \bibinfo{pages}{85--105}.
%Type = Article
\bibitem[{Dia and Javanshour(2017)}]{dia2017autonomous}
\bibinfo{author}{Dia, H.}, \bibinfo{author}{Javanshour, F.},
  \bibinfo{year}{2017}.
\newblock \bibinfo{title}{Autonomous shared mobility-on-demand: Melbourne pilot
  simulation study}.
\newblock \bibinfo{journal}{Transportation Research Procedia}
  \bibinfo{volume}{22}, \bibinfo{pages}{285--296}.
%Type = Incollection
\bibitem[{Fagiolo et~al.(2019)Fagiolo, Guerini, Lamperti, Moneta and
  Roventini}]{fagiolo2019validation}
\bibinfo{author}{Fagiolo, G.}, \bibinfo{author}{Guerini, M.},
  \bibinfo{author}{Lamperti, F.}, \bibinfo{author}{Moneta, A.},
  \bibinfo{author}{Roventini, A.}, \bibinfo{year}{2019}.
\newblock \bibinfo{title}{Validation of agent-based models in economics and
  finance}, in: \bibinfo{booktitle}{Computer Simulation Validation}.
  \bibinfo{publisher}{Springer}, pp. \bibinfo{pages}{763--787}.
%Type = Article
\bibitem[{Fagiolo and Roventini(2017)}]{fagiolo2017}
\bibinfo{author}{Fagiolo, G.}, \bibinfo{author}{Roventini, A.},
  \bibinfo{year}{2017}.
\newblock \bibinfo{title}{{Macroeconomic Policy in DSGE and Agent-Based Models
  Redux: New Developments and Challenges Ahead}}.
\newblock \bibinfo{journal}{Journal of Artificial Societies and Social
  Simulation} \bibinfo{volume}{20}, \bibinfo{pages}{1}.
\newblock \URLprefix \url{http://jasss.soc.surrey.ac.uk/20/1/1.html},
  \DOIprefix\doi{10.18564/jasss.3280}.
%Type = Article
\bibitem[{Fagnant and Kockelman(2013)}]{fagnant2013travel}
\bibinfo{author}{Fagnant, D.J.}, \bibinfo{author}{Kockelman, K.M.},
  \bibinfo{year}{2013}.
\newblock \bibinfo{title}{The travel and environmental implications of shared
  autonomous vehicles, using agent-based model scenarios}.
\newblock \bibinfo{journal}{Transportation Research Part C: Emerging
  Technologies} \bibinfo{volume}{40}, \bibinfo{pages}{1--13}.
%Type = Article
\bibitem[{Fagnant and Kockelman(2018)}]{fagnant2018dynamic}
\bibinfo{author}{Fagnant, D.J.}, \bibinfo{author}{Kockelman, K.M.},
  \bibinfo{year}{2018}.
\newblock \bibinfo{title}{Dynamic ride-sharing and fleet sizing for a system of
  shared autonomous vehicles in austin, texas}.
\newblock \bibinfo{journal}{Transportation} \bibinfo{volume}{45},
  \bibinfo{pages}{143--158}.
%Type = Article
\bibitem[{Fagnant et~al.(2015)Fagnant, Kockelman and
  Bansal}]{fagnant2015operations}
\bibinfo{author}{Fagnant, D.J.}, \bibinfo{author}{Kockelman, K.M.},
  \bibinfo{author}{Bansal, P.}, \bibinfo{year}{2015}.
\newblock \bibinfo{title}{Operations of shared autonomous vehicle fleet for
  austin, texas, market}.
\newblock \bibinfo{journal}{Transportation Research Record}
  \bibinfo{volume}{2563}, \bibinfo{pages}{98--106}.
%Type = Incollection
\bibitem[{Fellendorf and Vortisch(2010)}]{fellendorf2010microscopic}
\bibinfo{author}{Fellendorf, M.}, \bibinfo{author}{Vortisch, P.},
  \bibinfo{year}{2010}.
\newblock \bibinfo{title}{Microscopic traffic flow simulator vissim}, in:
  \bibinfo{booktitle}{Fundamentals of traffic simulation}.
  \bibinfo{publisher}{Springer}, pp. \bibinfo{pages}{63--93}.
%Type = Article
\bibitem[{Ferguson et~al.(2020)Ferguson, Laydon, Nedjati~Gilani, Imai, Ainslie,
  Baguelin, Bhatia, Boonyasiri, Cucunuba~Perez, Cuomo-Dannenburg
  et~al.}]{ferguson2020report}
\bibinfo{author}{Ferguson, N.}, \bibinfo{author}{Laydon, D.},
  \bibinfo{author}{Nedjati~Gilani, G.}, \bibinfo{author}{Imai, N.},
  \bibinfo{author}{Ainslie, K.}, \bibinfo{author}{Baguelin, M.},
  \bibinfo{author}{Bhatia, S.}, \bibinfo{author}{Boonyasiri, A.},
  \bibinfo{author}{Cucunuba~Perez, Z.}, \bibinfo{author}{Cuomo-Dannenburg, G.},
  et~al., \bibinfo{year}{2020}.
\newblock \bibinfo{title}{Report 9: Impact of non-pharmaceutical interventions
  (npis) to reduce covid19 mortality and healthcare demand} .
%Type = Article
\bibitem[{Haldane(2016)}]{haldane2016dappled}
\bibinfo{author}{Haldane, A.}, \bibinfo{year}{2016}.
\newblock \bibinfo{title}{The dappled world}.
\newblock \bibinfo{journal}{Shackle Biennial Memorial Lecture} .
%Type = Article
\bibitem[{H{\"o}rl(2017)}]{horl2017agent}
\bibinfo{author}{H{\"o}rl, S.}, \bibinfo{year}{2017}.
\newblock \bibinfo{title}{Agent-based simulation of autonomous taxi services
  with dynamic demand responses}.
\newblock \bibinfo{journal}{Procedia Computer Science} \bibinfo{volume}{109},
  \bibinfo{pages}{899--904}.
%Type = Inproceedings
\bibitem[{Iglesias et~al.(2018)Iglesias, Rossi, Wang, Hallac, Leskovec and
  Pavone}]{iglesias2018data}
\bibinfo{author}{Iglesias, R.}, \bibinfo{author}{Rossi, F.},
  \bibinfo{author}{Wang, K.}, \bibinfo{author}{Hallac, D.},
  \bibinfo{author}{Leskovec, J.}, \bibinfo{author}{Pavone, M.},
  \bibinfo{year}{2018}.
\newblock \bibinfo{title}{Data-driven model predictive control of autonomous
  mobility-on-demand systems}, in: \bibinfo{booktitle}{2018 IEEE International
  Conference on Robotics and Automation (ICRA)}, \bibinfo{organization}{IEEE}.
  pp. \bibinfo{pages}{1--7}.
%Type = Article
\bibitem[{Iglesias et~al.(2019)Iglesias, Rossi, Zhang and
  Pavone}]{iglesias2019bcmp}
\bibinfo{author}{Iglesias, R.}, \bibinfo{author}{Rossi, F.},
  \bibinfo{author}{Zhang, R.}, \bibinfo{author}{Pavone, M.},
  \bibinfo{year}{2019}.
\newblock \bibinfo{title}{A bcmp network approach to modeling and controlling
  autonomous mobility-on-demand systems}.
\newblock \bibinfo{journal}{The International Journal of Robotics Research}
  \bibinfo{volume}{38}, \bibinfo{pages}{357--374}.
%Type = Article
\bibitem[{James et~al.(2018)James, Lam and Lu}]{james2018double}
\bibinfo{author}{James, J.}, \bibinfo{author}{Lam, A.Y.}, \bibinfo{author}{Lu,
  Z.}, \bibinfo{year}{2018}.
\newblock \bibinfo{title}{Double auction-based pricing mechanism for autonomous
  vehicle public transportation system}.
\newblock \bibinfo{journal}{IEEE Transactions on Intelligent Vehicles}
  \bibinfo{volume}{3}, \bibinfo{pages}{151--162}.
%Type = Article
\bibitem[{Karamanis et~al.(2020)Karamanis, Anastasiadis, Angeloudis and
  Stettler}]{karamanis2020assignment}
\bibinfo{author}{Karamanis, R.}, \bibinfo{author}{Anastasiadis, E.},
  \bibinfo{author}{Angeloudis, P.}, \bibinfo{author}{Stettler, M.},
  \bibinfo{year}{2020}.
\newblock \bibinfo{title}{Assignment and pricing of shared rides in
  ride-sourcing using combinatorial double auctions}.
\newblock \bibinfo{journal}{IEEE Transactions on Intelligent Transportation
  Systems} .
%Type = Article
\bibitem[{Kleinrock(1975)}]{kleinrock1975queueing}
\bibinfo{author}{Kleinrock, L.}, \bibinfo{year}{1975}.
\newblock \bibinfo{title}{Queueing systems. volume i: theory} .
%Type = Article
\bibitem[{Korolko et~al.(2018)Korolko, Woodard, Yan and
  Zhu}]{korolko2018dynamic}
\bibinfo{author}{Korolko, N.}, \bibinfo{author}{Woodard, D.},
  \bibinfo{author}{Yan, C.}, \bibinfo{author}{Zhu, H.}, \bibinfo{year}{2018}.
\newblock \bibinfo{title}{Dynamic pricing and matching in ride-hailing
  platforms}.
\newblock \bibinfo{journal}{Available at SSRN} .
%Type = Article
\bibitem[{Levin et~al.(2017)Levin, Kockelman, Boyles and Li}]{levin2017general}
\bibinfo{author}{Levin, M.W.}, \bibinfo{author}{Kockelman, K.M.},
  \bibinfo{author}{Boyles, S.D.}, \bibinfo{author}{Li, T.},
  \bibinfo{year}{2017}.
\newblock \bibinfo{title}{A general framework for modeling shared autonomous
  vehicles with dynamic network-loading and dynamic ride-sharing application}.
\newblock \bibinfo{journal}{Computers, Environment and Urban Systems}
  \bibinfo{volume}{64}, \bibinfo{pages}{373--383}.
%Type = Inproceedings
\bibitem[{Lin et~al.(2018)Lin, Zhao, Xu and Zhou}]{lin2018efficient}
\bibinfo{author}{Lin, K.}, \bibinfo{author}{Zhao, R.}, \bibinfo{author}{Xu,
  Z.}, \bibinfo{author}{Zhou, J.}, \bibinfo{year}{2018}.
\newblock \bibinfo{title}{Efficient large-scale fleet management via
  multi-agent deep reinforcement learning}, in: \bibinfo{booktitle}{Proceedings
  of the 24th ACM SIGKDD International Conference on Knowledge Discovery \&
  Data Mining}, pp. \bibinfo{pages}{1774--1783}.
%Type = Article
\bibitem[{Liu et~al.(2017)Liu, Kockelman, Boesch and Ciari}]{liu2017tracking}
\bibinfo{author}{Liu, J.}, \bibinfo{author}{Kockelman, K.M.},
  \bibinfo{author}{Boesch, P.M.}, \bibinfo{author}{Ciari, F.},
  \bibinfo{year}{2017}.
\newblock \bibinfo{title}{Tracking a system of shared autonomous vehicles
  across the austin, texas network using agent-based simulation}.
\newblock \bibinfo{journal}{Transportation} \bibinfo{volume}{44},
  \bibinfo{pages}{1261--1278}.
%Type = Article
\bibitem[{Macal and North(2010)}]{doi:10.1057/jos.2010.3}
\bibinfo{author}{Macal, C.M.}, \bibinfo{author}{North, M.J.},
  \bibinfo{year}{2010}.
\newblock \bibinfo{title}{{Tutorial on agent-based modelling and simulation}}.
\newblock \bibinfo{journal}{Journal of Simulation} \bibinfo{volume}{4},
  \bibinfo{pages}{151--162}.
\newblock \URLprefix \url{https://doi.org/10.1057/jos.2010.3},
  \DOIprefix\doi{10.1057/jos.2010.3}.
%Type = Article
\bibitem[{Maciejewski and Bischoff(2016)}]{maciejewski2016congestion}
\bibinfo{author}{Maciejewski, M.}, \bibinfo{author}{Bischoff, J.},
  \bibinfo{year}{2016}.
\newblock \bibinfo{title}{Congestion effects of autonomous taxi fleets} .
%Type = Article
\bibitem[{Maciejewski et~al.(2016)Maciejewski, Salanova, Bischoff and
  Estrada}]{maciejewski2016large}
\bibinfo{author}{Maciejewski, M.}, \bibinfo{author}{Salanova, J.M.},
  \bibinfo{author}{Bischoff, J.}, \bibinfo{author}{Estrada, M.},
  \bibinfo{year}{2016}.
\newblock \bibinfo{title}{Large-scale microscopic simulation of taxi services.
  berlin and barcelona case studies}.
\newblock \bibinfo{journal}{Journal of Ambient Intelligence and Humanized
  Computing} \bibinfo{volume}{7}, \bibinfo{pages}{385--393}.
%Type = Article
\bibitem[{Mankiw(2006)}]{mankiw2006macroeconomist}
\bibinfo{author}{Mankiw, N.G.}, \bibinfo{year}{2006}.
\newblock \bibinfo{title}{The macroeconomist as scientist and engineer}.
\newblock \bibinfo{journal}{Journal of Economic Perspectives}
  \bibinfo{volume}{20}, \bibinfo{pages}{29--46}.
%Type = Inproceedings
\bibitem[{Martinez and Crist(2015)}]{martinez2015urban}
\bibinfo{author}{Martinez, L.}, \bibinfo{author}{Crist, P.},
  \bibinfo{year}{2015}.
\newblock \bibinfo{title}{Urban mobility system upgrade--how shared
  self-driving cars could change city traffic}, in:
  \bibinfo{booktitle}{International Transport Forum, Paris}.
%Type = Inproceedings
\bibitem[{Martinez and Viegas(2016)}]{martinez2016shared}
\bibinfo{author}{Martinez, L.}, \bibinfo{author}{Viegas, J.},
  \bibinfo{year}{2016}.
\newblock \bibinfo{title}{Shared mobility: innovation for livable cities}, in:
  \bibinfo{booktitle}{International Transport Forum}.
%Type = Article
\bibitem[{McNally(2000)}]{mcnally2000four}
\bibinfo{author}{McNally, M.G.}, \bibinfo{year}{2000}.
\newblock \bibinfo{title}{The four step model}.
\newblock \bibinfo{journal}{Handbook of transport modelling}
  \bibinfo{volume}{1}, \bibinfo{pages}{35--41}.
%Type = Inproceedings
\bibitem[{Nagel et~al.(1999)Nagel, Beckman and Barrett}]{nagel1999transims}
\bibinfo{author}{Nagel, K.}, \bibinfo{author}{Beckman, R.J.},
  \bibinfo{author}{Barrett, C.L.}, \bibinfo{year}{1999}.
\newblock \bibinfo{title}{Transims for urban planning}, in:
  \bibinfo{booktitle}{6th international conference on computers in urban
  planning and urban management, Venice, Italy}.
%Type = Article
\bibitem[{Narayanan et~al.(2020)Narayanan, Chaniotakis and
  Antoniou}]{narayanan2020shared}
\bibinfo{author}{Narayanan, S.}, \bibinfo{author}{Chaniotakis, E.},
  \bibinfo{author}{Antoniou, C.}, \bibinfo{year}{2020}.
\newblock \bibinfo{title}{Shared autonomous vehicle services: A comprehensive
  review}.
\newblock \bibinfo{journal}{Transportation Research Part C: Emerging
  Technologies} \bibinfo{volume}{111}, \bibinfo{pages}{255--293}.
%Type = Article
\bibitem[{Qiu et~al.(2018)Qiu, Li and Zhao}]{qiu2018dynamic}
\bibinfo{author}{Qiu, H.}, \bibinfo{author}{Li, R.}, \bibinfo{author}{Zhao,
  J.}, \bibinfo{year}{2018}.
\newblock \bibinfo{title}{Dynamic pricing in shared mobility on demand
  service}.
\newblock \bibinfo{journal}{arXiv:1802.03559} .
%Type = Book
\bibitem[{Ross(2006)}]{ross2006introduction}
\bibinfo{author}{Ross, S.M.}, \bibinfo{year}{2006}.
\newblock \bibinfo{title}{Introduction to Probability Models, ISE}.
\newblock \bibinfo{publisher}{Academic press}.
%Type = Misc
\bibitem[{{San Francisco County Transportation
  Authority}(2020)}]{SanFranciscoCountyTransportationAuthority2020}
\bibinfo{author}{{San Francisco County Transportation Authority}},
  \bibinfo{year}{2020}.
\newblock \bibinfo{title}{{TNCs Today}}.
\newblock \URLprefix \url{https://www.sfcta.org/projects}.
%Type = Article
\bibitem[{Santi et~al.(2014)Santi, Resta, Szell, Sobolevsky, Strogatz and
  Ratti}]{santi2014quantifying}
\bibinfo{author}{Santi, P.}, \bibinfo{author}{Resta, G.},
  \bibinfo{author}{Szell, M.}, \bibinfo{author}{Sobolevsky, S.},
  \bibinfo{author}{Strogatz, S.H.}, \bibinfo{author}{Ratti, C.},
  \bibinfo{year}{2014}.
\newblock \bibinfo{title}{Quantifying the benefits of vehicle pooling with
  shareability networks}.
\newblock \bibinfo{journal}{Proceedings of the National Academy of Sciences}
  \bibinfo{volume}{111}, \bibinfo{pages}{13290--13294}.
%Type = Misc
\bibitem[{Schneider(2020)}]{Schneider2020}
\bibinfo{author}{Schneider, T.W.}, \bibinfo{year}{2020}.
\newblock \bibinfo{title}{{Analyzing 1.1 Billion NYC Taxi and Uber Trips, with
  a Vengeance}}.
\newblock \URLprefix \url{https://toddwschneider.com}.
%Type = Inproceedings
\bibitem[{Shen and Lopes(2015)}]{shen2015managing}
\bibinfo{author}{Shen, W.}, \bibinfo{author}{Lopes, C.}, \bibinfo{year}{2015}.
\newblock \bibinfo{title}{Managing autonomous mobility on demand systems for
  better passenger experience}, in: \bibinfo{booktitle}{International
  conference on principles and practice of multi-agent systems},
  \bibinfo{organization}{Springer}. pp. \bibinfo{pages}{20--35}.
%Type = Article
\bibitem[{Simon(1955)}]{simon1955behavioral}
\bibinfo{author}{Simon, H.A.}, \bibinfo{year}{1955}.
\newblock \bibinfo{title}{A behavioral model of rational choice}.
\newblock \bibinfo{journal}{The quarterly journal of economics}
  \bibinfo{volume}{69}, \bibinfo{pages}{99--118}.
%Type = Misc
\bibitem[{TLC(2019)}]{TLC2019}
\bibinfo{author}{TLC}, \bibinfo{year}{2019}.
\newblock \bibinfo{title}{{New York City Taxi Trip Data}}.
\newblock \URLprefix \url{https://www1.nyc.gov/site/tlc}.
%Type = Article
\bibitem[{Turrell(2016)}]{turrell2016agent}
\bibinfo{author}{Turrell, A.}, \bibinfo{year}{2016}.
\newblock \bibinfo{title}{Agent-based models: understanding the economy from
  the bottom up}.
\newblock \bibinfo{journal}{Bank of England Quarterly Bulletin} ,
  \bibinfo{pages}{Q4}.
%Type = Article
\bibitem[{Von~Neumann et~al.(1966)Von~Neumann, Burks et~al.}]{von1966theory}
\bibinfo{author}{Von~Neumann, J.}, \bibinfo{author}{Burks, A.W.}, et~al.,
  \bibinfo{year}{1966}.
\newblock \bibinfo{title}{Theory of self-reproducing automata}.
\newblock \bibinfo{journal}{IEEE Transactions on Neural Networks}
  \bibinfo{volume}{5}, \bibinfo{pages}{3--14}.
%Type = Inproceedings
\bibitem[{Wallar et~al.(2018)Wallar, Van Der~Zee, Alonso-Mora and
  Rus}]{wallar2018vehicle}
\bibinfo{author}{Wallar, A.}, \bibinfo{author}{Van Der~Zee, M.},
  \bibinfo{author}{Alonso-Mora, J.}, \bibinfo{author}{Rus, D.},
  \bibinfo{year}{2018}.
\newblock \bibinfo{title}{Vehicle rebalancing for mobility-on-demand systems
  with ride-sharing}, in: \bibinfo{booktitle}{2018 IEEE/RSJ International
  Conference on Intelligent Robots and Systems (IROS)},
  \bibinfo{organization}{IEEE}. pp. \bibinfo{pages}{4539--4546}.
%Type = Article
\bibitem[{Wang and Yang(2019)}]{Wang2019}
\bibinfo{author}{Wang, H.}, \bibinfo{author}{Yang, H.}, \bibinfo{year}{2019}.
\newblock \bibinfo{title}{{Ridesourcing systems : A framework and review}}.
\newblock \bibinfo{journal}{Transportation Research Part B}
  \bibinfo{volume}{129}, \bibinfo{pages}{122--155}.
\newblock \URLprefix \url{https://doi.org/10.1016/j.trb.2019.07.009},
  \DOIprefix\doi{10.1016/j.trb.2019.07.009}.
%Type = Inproceedings
\bibitem[{Wen et~al.(2017)Wen, Zhao and Jaillet}]{wen2017rebalancing}
\bibinfo{author}{Wen, J.}, \bibinfo{author}{Zhao, J.},
  \bibinfo{author}{Jaillet, P.}, \bibinfo{year}{2017}.
\newblock \bibinfo{title}{Rebalancing shared mobility-on-demand systems: A
  reinforcement learning approach}, in: \bibinfo{booktitle}{2017 IEEE 20th
  International Conference on Intelligent Transportation Systems (ITSC)},
  \bibinfo{organization}{IEEE}. pp. \bibinfo{pages}{220--225}.
%Type = Article
\bibitem[{Yang et~al.(2002)Yang, Wong and Wong}]{Yang2002}
\bibinfo{author}{Yang, H.}, \bibinfo{author}{Wong, S.C.},
  \bibinfo{author}{Wong, K.I.}, \bibinfo{year}{2002}.
\newblock \bibinfo{title}{{Demand-supply equilibrium of taxi services in a
  network under competition and regulation}}.
\newblock \bibinfo{journal}{Transportation Research Part B: Methodological}
  \bibinfo{volume}{36}, \bibinfo{pages}{799--819}.
\newblock \DOIprefix\doi{10.1016/S0191-2615(01)00031-5}.
%Type = Article
\bibitem[{Zha et~al.(2018)Zha, Yin and Xu}]{zha2018geometric}
\bibinfo{author}{Zha, L.}, \bibinfo{author}{Yin, Y.}, \bibinfo{author}{Xu, Z.},
  \bibinfo{year}{2018}.
\newblock \bibinfo{title}{Geometric matching and spatial pricing in
  ride-sourcing markets}.
\newblock \bibinfo{journal}{Transportation Research Part C: Emerging
  Technologies} \bibinfo{volume}{92}, \bibinfo{pages}{58--75}.
%Type = Article
\bibitem[{Zhang and Pavone(2016)}]{zhang2016control}
\bibinfo{author}{Zhang, R.}, \bibinfo{author}{Pavone, M.},
  \bibinfo{year}{2016}.
\newblock \bibinfo{title}{Control of robotic mobility-on-demand systems: a
  queueing-theoretical perspective}.
\newblock \bibinfo{journal}{The International Journal of Robotics Research}
  \bibinfo{volume}{35}, \bibinfo{pages}{186--203}.

\end{thebibliography}

\end{document}